\documentclass{aa}

\usepackage{graphicx}
\usepackage{txfonts}
\usepackage{xcolor}
\usepackage{amsmath}
\usepackage{amssymb}
\usepackage{ulem} 

\usepackage[colorlinks=true,citecolor=blue,urlcolor=blue]{hyperref}

\usepackage{txfonts}
\usepackage{booktabs}
\usepackage[utf8]{inputenc}
\begin{document}


   \title{The Galactic inner spiral arms revealed by the {\it Gaia} ESO Survey chemical abundances}
   \subtitle{Metallicity and [Mg/Fe] ratios}

 \author{
C. Viscasillas V{\'a}zquez\inst{\ref{vilnius}}\and 
L. Magrini\inst{\ref{oaa}} \and 
E. Spitoni\inst{\ref{oat}, \ref{ifpu}  } \and
G. Cescutti\inst{\ref{oat}, \ref{trieste_fis}, \ref{trieste_infn}} \and
G. Tautvai{\v s}ien{\. e}\inst{\ref{vilnius}} \and 
A. Vasini\inst{\ref{clap},\ref{oat} } \and
S. Randich \inst{\ref{oaa}}  \and
G.G. Sacco \inst{\ref{oaa}}}

\institute{
Institute of Theoretical Physics and Astronomy, Vilnius University, Sauletekio av. 3, 10257 Vilnius, Lithuania. \label{vilnius}
\and 
INAF - Osservatorio Astrofisico di Arcetri, Largo E. Fermi 5, 50125, Firenze, Italy.  \label{oaa} 
\and
INAF, Osservatorio Astronomico di Trieste, Via Tiepolo 11, I-34143 Trieste, Italy. \label{oat} 
\and IFPU, Institute for Fundamental Physics of the Universe, Via Beirut 2, I-34151 Trieste, Italy. \label{ifpu} 
\and Dipartimento di Fisica, Sezione di Astronomia, Università di Trieste, Via G. B. Tiepolo 11, 34143 Trieste, Italy. \label{trieste_fis} 
\and INFN, Sezione di Trieste, Via A. Valerio 2, 34127 Trieste, Italy. \label{trieste_infn} 
\and Como Lake centre for AstroPhysics (CLAP), DiSAT, Università dell’Insubria, via Valleggio 11, 22100 Como, Italy. \label{clap}
}

   \date{Received 21 February 2025 / Accepted 24 April 2025 }

 
  \abstract
   {The spiral structure of the Milky Way has traditionally been mapped using stellar density, kinematics, and gas distribution. However, chemical abundances—especially in the inner Galaxy—remain a relatively under-explored tracer, offering valuable insights into its formation and evolution. Recent observational advances, such as {\it Gaia} DR3 GSP-Spec, have highlighted the potential of chemical abundances in tracing and revealing the structure of spiral arms.}
   {Building on these studies, we aim to trace the Milky Way's inner spiral arms using chemical abundance data from the {\it Gaia}-ESO Survey (GES). By mapping over-densities in [Fe/H] and [Mg/Fe], we seek to identify spiral arms in both radial and vertical planes, detect substructures, and compare our results with recent Galactic chemical evolution models.}
   {We used chemical abundance data from the {\it Gaia}-ESO Survey to create spatial maps of [Fe/H], [Mg/H], and [Mg/Fe] excess across the Galactic inner disc. The maps were analysed to detect over-densities associated with known spiral arms (Drimmel et al. 2024). We compared our results with the spiral arm models proposed by Spitoni et al. (2023) and Barbillon et al. (2024).}
   {For the first time, the inner spiral arms were revealed using chemical abundance patterns. We detected [Fe/H] enhancements and [Mg/Fe] under-abundances that consistently trace the Scutum and Sagittarius arms. A connecting spur between these arms is observed in the [Mg/H] plane.  The alignment between our observations and the results of our 2D chemical evolution models reinforces the significance of spiral arm transits in driving both azimuthal and radial variations in chemical abundances.
   }
   {Our results confirm that spiral arms can be traced using stellar chemical abundances with GES data, providing a new perspective on the structure of the inner Galaxy. The consistency between enhanced [Fe/H] and lower [Mg/Fe] ratios, as observed in previous studies, further strengthens the reliability of our findings. The observed spur, bifurcation, and vertical substructures align well with recent models and studies, indicating that chemical maps can significantly contribute to our understanding of Galactic spiral arms.}

   \keywords{Galaxy: disc – Galaxy: evolution – Galaxy: structure – Galaxy: abundances – Galaxy: stellar content
               }

   \maketitle
%

\section{Introduction}

Spiral arms play a crucial role in shaping the chemical evolution of disc galaxies, and observations in external galaxies have revealed enhanced metallicity within the arms compared to the inter-arm regions \citep[see, e.g.][]{Sanchez_Menguiano_2020}. Moreover, the spiral structure can influence metallicity not only on a local scale but also across larger distances \citep[see, e.g][]{Lepine2011}.

Recent advancements in astrometric and photometric data from the {\it Gaia} mission have significantly enhanced our under-standing of the Milky Way’s spiral structure and its chemical evolution. \citet{Poggio2021} utilized the {\it Gaia}  Early Data Release 3 (EDR3) \citep{GaiaCollaboration2021} to map the
density variations of young upper main sequence stars, open clusters (OC), and classical Cepheids within a few kiloparsecs of the Sun \citep[see, also][]{Ge_2024,Castro-Ginard2021,Lemasle2022}. Their findings revealed coherent large-scale arches corresponding to spiral arm segments, notably suggesting an extended Local Arm stretching at least 8 kpc from the Sun. Building on this, \citet{Poggio2022} utilized the {\it Gaia} Data Release 3 (DR3) \citep{GaiaCollaboration2023} and {\it Gaia} DR3 General Stellar Parametrizer Spectroscopy (GSP-Spec) \citep{RecioBlanco2023} to uncover chemical inhomogeneities in the Galactic disc, identifying metal-rich features aligned with spiral arms. These chemical signatures are particularly pronounced in younger stellar populations, underscoring the crucial role of the spiral arms in shaping the Galaxy's chemical landscape. Additionally, \citet{Barbillon2024} extended this work by examining azimuthal variations in [Ca/Fe] and [Mg/Fe], revealing distinct chemical signatures linked to the spiral arms, such as those associated with the Sagittarius-Carina and Local arms. Their analysis revealed [$\alpha$/Fe] depletions in young stars within spiral arms compared to inter-arm regions, attributed to enhanced iron production within the arms. The [Mg/Fe] ratio, or in general [$\alpha$/Fe],  is, indeed,  a valuable tracer of star formation history, as magnesium is predominantly produced by core-collapse supernovae (SNe II) on short timescales \citep[e.g][]{Chieffi2004ApJ...608..405C, Chieffi2013ApJ...764...21C, Higgins2023MNRAS.526..534H}, while iron has significant contributions from Type Ia supernovae (SNe Ia) on longer timescales \citep[see, e.g.][]{Nomoto2013ARA&A..51..457N}. Variations in the [Mg/Fe] ratio  across different galaxies or Galactic populations—such as the thin and thick discs, halo, and bulge—reflect differences in their star formation efficiencies and chemical evolution, with higher [Mg/Fe] ratios typically indicating more rapid star formation.

In addition to studies of the chemical composition of spiral arm regions, further works based on stellar density, stellar tracers of different ages, and kinematics have increased our knowledge of the spiral pattern, especially on a local scale \citep[see, e.g.][]{Rezaei_2018,   Khoperskov_2022, Lin_2022, Ge_2024, Widmark_2024}. 
For instance, \citet{Reid2019} compiled trigonometric parallaxes of molecular masers in high-mass star-forming regions, supporting a four-arm spiral model, whereas \citet{Lemasle2022} and \citet{Drimmel2024} utilized classical Cepheids  and  young supergiants to trace the spiral arms at even greater distances, deriving updated parameters for the Galactic warp.
Complementing these works, \citet{Alinder2024} analysed the two-armed phase spiral within the solar neighbourhood using {\it Gaia} DR3 data, uncovering its distinct rotational behaviour and chemical signature compared to the one-armed phase spiral. On the other side, recent studies revealed that nova systems, dominating $^{26}$Al production with $\sim$75\%, dampen the Milky Way's spiral arm pattern, highlighting their critical role in the Galaxy's chemical evolution alongside massive stars \citep{Vasini2025}. Together, these studies provide new insights into the Galaxy’s spiral architecture and the complex history of dynamical disturbances in its disc.
From a dynamical point of view, \citet{Palicio2023} utilized {\it Gaia} DR3 radial velocity data and radial actions to identify spiral-like features in the Galactic disc, revealing arc-like segments and high radial action ($J_R$) regions, such as one around $R_{\rm GC}$=10.5 kpc, potentially linked to the Outer Lindblad Resonance. Their work highlights the consistency between the dynamics of older stars and the spiral arms traced by younger populations, reinforcing the complex interplay between Galactic structure and stellar motion.\\
Complementing the observational insights, \citet{Spitoni2023} developed a set of advanced 2D chemical evolution models incorporating multiple spiral-arm patterns with varying pattern speeds. Those models successfully reproduce the azimuthal variations in the abundance gradients of elements like oxygen, iron, europium, and barium. They underscore the influence of distinct spiral modes on the distribution of chemically enriched materials in the disc, aligning well with the chemical inhomogeneities observed by \citet{Poggio2021}. 

In the present work, we took advantage of the spectroscopic data of the {\it Gaia}-ESO survey \citep[GES hereafter,][]{Randich2022, Gilmore2022} to investigate the azimuthal variations of metallicity and abundances, and their correlation with the spiral pattern. Compared to previous studies, the {\it Gaia}-ESO data allowed us to investigate the innermost regions of the disc, corresponding to the Scutum and Norma spiral  arms. Furthermore, the results can be compared with those of previous studies, given the good agreement of {\it Gaia} GSP-Spec with GES \citep{VanderSwaelmen2024}. Following the above, the paper is structured as follows: in Section \ref{sec:dataset}, we describe the dataset used in this study, focusing on the chemical abundance measurements provided by the {\it Gaia}-ESO Survey  and their spatial coverage across the Galactic disc. Section \ref{sec:maps} presents metallicity and abundance ratio excess maps. In Section \ref{sec:residuals}, we explore the residual abundance variations across Galactic longitude, highlighting key trends and their potential connection to the spiral structure of the inner Galaxy in comparison with recent models. Finally, Section \ref{sec:conclusions} summarizes our findings and discusses their implications for understanding the chemical and structural evolution of the Galactic disc.
\section{Dataset}
\label{sec:dataset}


\begin{figure}
  \resizebox{\hsize}{!}{\includegraphics{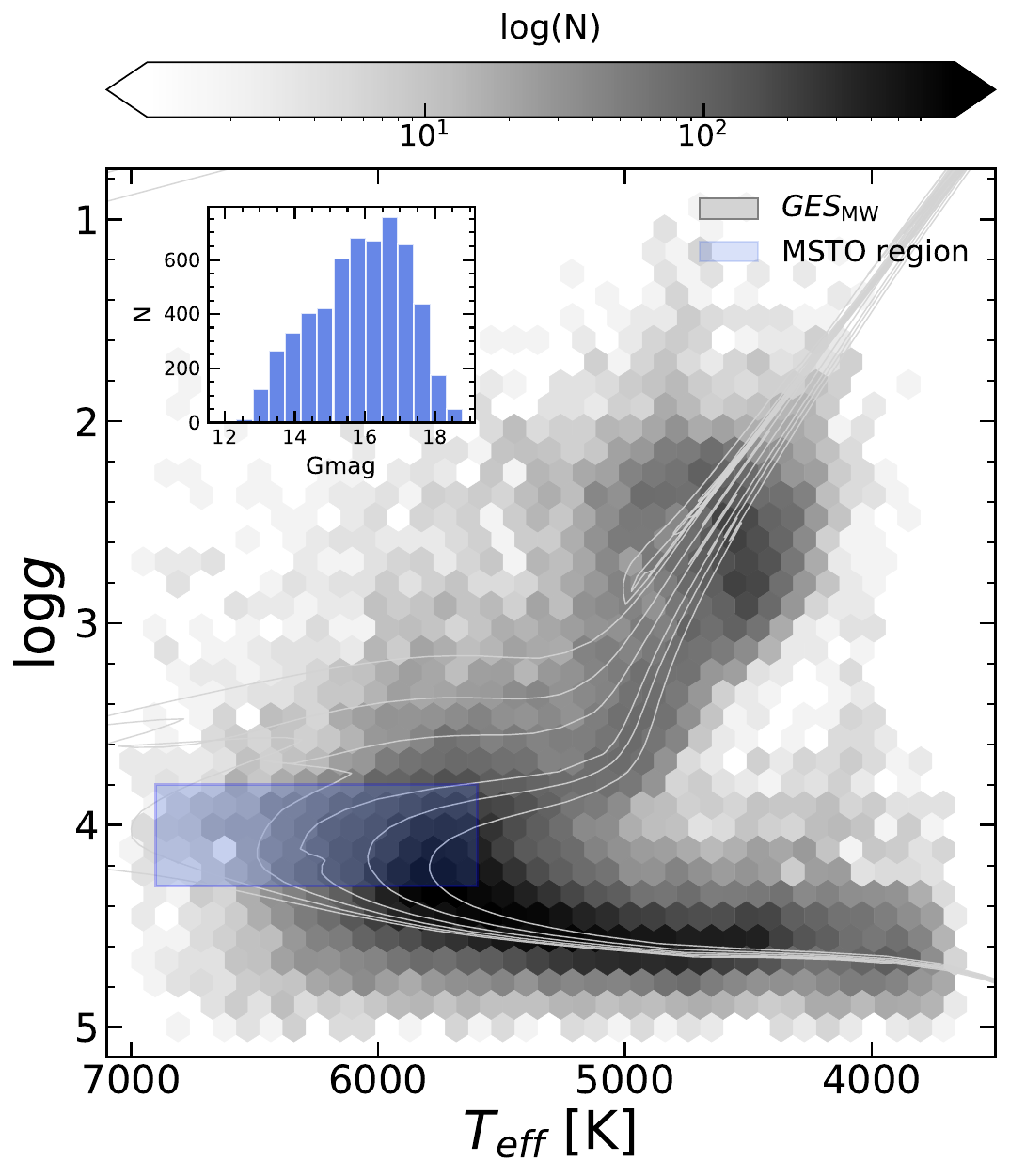}}
  \caption{Kiel diagram showing the distribution of \texttt{\textit{GE\_MW}} and \textit{GE\_MW\_BL} stars in the $T_{\rm eff}$-log$g$ plane. The background grayscale density plot represents the full sample of \texttt{\textit{GE\_MW}} and \textit{GE\_MW\_BL} stars, with a logarithmic binning. The blue-shaded rectangle highlights the region used to select main sequence turn-off (MSTO) stars. The inset panel shows the distribution of Gaia G magnitudes for the selected stars. In lightgray we show grids of PARSEC isochrones \citep{Bressan2012}, with log(age/yr) from 9.0 to 10.0 at solar metallicity.}
  \label{fig:kiel}
\end{figure}

\begin{figure}
  \resizebox{\hsize}{!}{\includegraphics{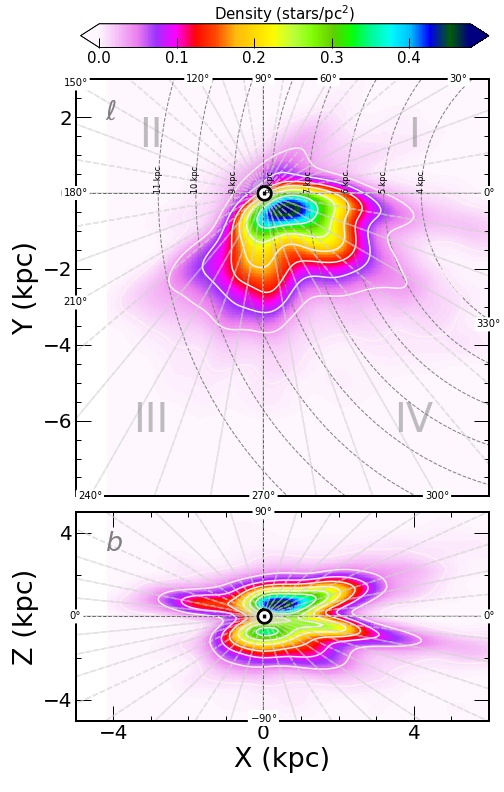}}
  \caption{Density plot of approximately $\sim$5,000 MSTO field stars with metallicity [Fe/H] > $-$0.5. The plot shows the Galactic coordinate system projected onto the XY plane. Lines of constant Galactic longitude ($\ell$) and latitude ($b$) are drawn at regular intervals, while concentric circles represent increasing Galactocentric distances ($R$). The four Galactic quadrants are also indicated, providing a reference for the spatial distribution of structures within the Milky Way. The colourbar shows the stellar density in physical units of number of stars per square parsec (stars/pc$^2$)}
  \label{fig:XYZ_density}
\end{figure}

Using the latest release of GES, which includes approximately 114,000 unique targets \citep{Hourihane2023A&A...676A.129H}, we initially selected around $\sim$63,000 stars in the Galactic fields, identified with \texttt{GES\_TYPE = \textit{GE\_MW}}, including approximately $\sim$6,000 stars located in the Galactic Bulge, classified as \texttt{GES\_TYPE = \textit{GE\_MW\_BL}}.
These stars were observed both with GIRAFFE at medium resolution ($R=20000$) and  UVES at high resolution ($R=47000$).  From this subset, we further narrowed down our selection including only  Main Sequence Turn-Off (MSTO) stars (see Figure \ref{fig:kiel}), following the criteria set in \citet{Viscasillas2023}, and retained only those with typical thin disc metallicity, [Fe/H] > $-$0.5, reducing the sample to around $\sim$5,000 field stars ($\sim$73\% of them were observed with Giraffe HR10 and/or HR21 setups and $\sim$27\% with UVES U580). This selection allows us to more accurately map the spiral structure, as the selected stars  are expected to be younger, more metal-rich, and predominantly located in the Galactic thin disc, the region where spiral structures are most prominent.  Approximately 66\% of them are located in the 4th Galactic quadrant ($270^\circ \leq \ell \leq 360^\circ$), 25\% in the 3th ($180^\circ \leq \ell \leq 270^\circ$) and 73\% are located in the combined 1st ($0^\circ \leq \ell \leq 90^\circ$) and 4th Galactic quadrants ($270^\circ \leq \ell \leq 360^\circ$). In Figure \ref{fig:XYZ_density} we show both the distribution of our sample on the Galactic plane and on the vertical direction, above and below the plane. 
The distribution of our sample of field stars enables us to map not only the Sagittarius arm but also the inner arms, such as the Scutum arm and even up to the Norma arm, complementing other studies that focus mainly on the outer regions of the Galaxy \citep{Poggio2021, Poggio2022, Spitoni2023, Barbillon2024}. The sub-panel in Figure~\ref{fig:kiel} also displays the G magnitude distribution of our selected sample: the brightest stars were observed with UVES, while the faintest ones were reached with GIRAFFE. 
We computed ages of MSTO field stars using the {\sc aussieq2} code{\footnote{{https://github.com/spinastro/aussieq2}}, based on the original {\sc q2} code from \citet{ramirez2014A&A...572A..48R}, which derives stellar ages and masses comparing stellar parameters with a grid of isochrones.

Additionally, we incorporated a sample of 62 OCs as defined in \citet{Viscasillas2022} and \citet{Magrini2023}. Heliocentric Cartesian coordinates of the field sample were taken from \citet{Candebat2024} and for the OCs, they were computed in this work using the {\sc Galpy} code \citep{bovy15}.



\section{Metallicity and abundance ratio excess maps}
\label{sec:maps}

The distribution of stars along the line of sight from the Sun, together with the limited sample size, makes it challenging to discern the spiral arms from stellar density alone. To address this, we constructed smooth, continuous maps of [Fe/H], [Mg/Fe], and [Mg/H] excesses in two projections (XY and XZ), computed using the methodology detailed below.
First, we interpolated the original dataset using the \texttt{griddata} function from {\sc SciPy} \citep{Virtanen2020}. Then, we generated uniform grids for each projection and applied linear interpolation to estimate the abundance values across these grids. To reduce noise and emphasize spatial patterns, we applied a \texttt{Gaussian filter} from SciPy, ensuring the resulting maps were clearer and more representative of large-scale trends \citep[see a similar approach in e.g.][]{Alinder2024}. The overlaid spiral arm model is based on the framework proposed by \citet{Drimmel2024}. Finally, to highlight localized variations, we computed excess maps for [Fe/H], [Mg/Fe] and [Mg/H] following the approach described by \citet{Poggio2022}. For each abundance ratio, we smoothed the data at two scales using Gaussian filters: a finer scale ($\sigma = 7$) to capture local variations and a broader scale ($\sigma = 21$) to represent global trends, where $\sigma$ is the standard deviation of the Gaussian kernel. By subtracting the global trend map from the locally smoothed map, we isolated regions of significant deviations, effectively identifying areas of localized enrichment or depletion in [Fe/H], [Mg/Fe] and [Mg/H]. This method inherently reduces the spatial coverage of the excess map, as smooth large-scale structures are removed by construction.

In the case of [Fe/H], these excess maps reveal subtle chemical inhomogeneities within the Galactic disc (see Figure \ref{fig:XYZ_FEH_excess_clusters}). Based on their positions, they appear to correspond to the Sagittarius, Scutum, and Norma arms. On the maps, they appear with a width consistent with the typical estimated range of $\sim$500–2000 pc \citep[see, e.g.,][]{Valle2014, Reid2019}. In contrast, when examining the same figure in terms of [Mg/Fe] excess (see Figure \ref{fig:XYZ_MGFE_excess}), we find that the regions associated with the spiral arms appear depleted in [Mg/Fe]. This depletion suggests that while these regions exhibit localized enrichment in iron, they are relatively poorer in magnesium compared to their surroundings. This effect is particularly evident in the vertical view of the referenced figure. Our findings align with those of \citet{Barbillon2024}, which indicate enhanced iron production in spiral arms. Similarly, the magnesium excess maps (see Figure \ref{fig:XYZ_MGH_excess_clusters}) reveal regions with significant deviations in [Mg/H], offering insights into the differential enrichment processes that may be influencing various areas.

The figures also reveal a potential substructure or bifurcation of the Scutum Arm at a Galactic longitude of approximately $\ell$ = 320$^\circ$ and at Y = $\sim$$-$3 kpc, in the {\sc IV} quadrant. This appears to produce a splitting into an inner and outer sub-arm. These types of extensions are frequently found in observed multi-armed galaxies. Indeed, the Scutum Arm together with the Perseus Arm are considered the primary arms of the Galaxy, while the Norma and Sagittarius arms are secondary \citep[e.g.][]{Valle2016}. This splitting of one main arm into two seems to align with the findings of \citet{Xu2023}, who used masers and high-mass star-forming regions to identify similar structures. However, the authors of that study consider two major spiral arms, the Norma and Perseus Arms. In that case, Scutum would be a fork of the Norma Arm. Models also suggest that in a initially two-arm structure near the Galactic Center, each inner arm undergoes a bifurcation, resulting in two additional gaseous arms beyond \citep[see e.g.][]{Perez_Villegas_2015}. This bifurcation at a galactocentric distance of $\sim$5 kpc also appears to be seen in Model 3 of \citet{Vislosky2024}, and the Scutum–Centaurus spiral arm appears connected to the bar’s near side. Furthermore, when we look at the maps in the [Mg/H] plane (see Figure \ref{fig:XYZ_MGH_excess_clusters}), we can see branches or bridges that go from one spiral arm to another. Particularly, in addition to distinguishing the arms at low Y coordinates, we can see what appears to be a connection between Sagittarius and Scutum at approximately $\ell$ = 280$^\circ$. Indeed, branches, spurs, and/or feathers are frequently observed in Sb-Sc galaxies, forming periodic structures associated with dust lanes, gas density peaks, and gravitational instabilities \citep[see e.g.][]{Elmegreen1980,Dobbs2006,LaVigne2006,Hou2021} for a review. 
Figure \ref{fig:XYZ_FEH_excess_clusters} also appears to show part of an interior substructure at $\ell$ = 330-340$^\circ$ and at a Galactocentric distance of about $\sim$4 kpc, which could correspond to the outermost part of Norma or even to the recently discovered feather by \citet{Veena2021}, called the Gangotri wave. This feather is located at the same coordinates (see Figure 4 of the aforementioned work). 
Finally, we note that the vertical view in our study suggests that the innermost arm or structure extends slightly below the Galactic plane, while the other inner arm remains closer to it, and the outermost structure lies slightly above. Additionally, there appears to be a gradual increase in the height of the spiral arms above the Galactic plane from the inside out, which may be consistent with the model of a warped Galaxy.

\begin{figure} \resizebox{\hsize}{!}{\includegraphics{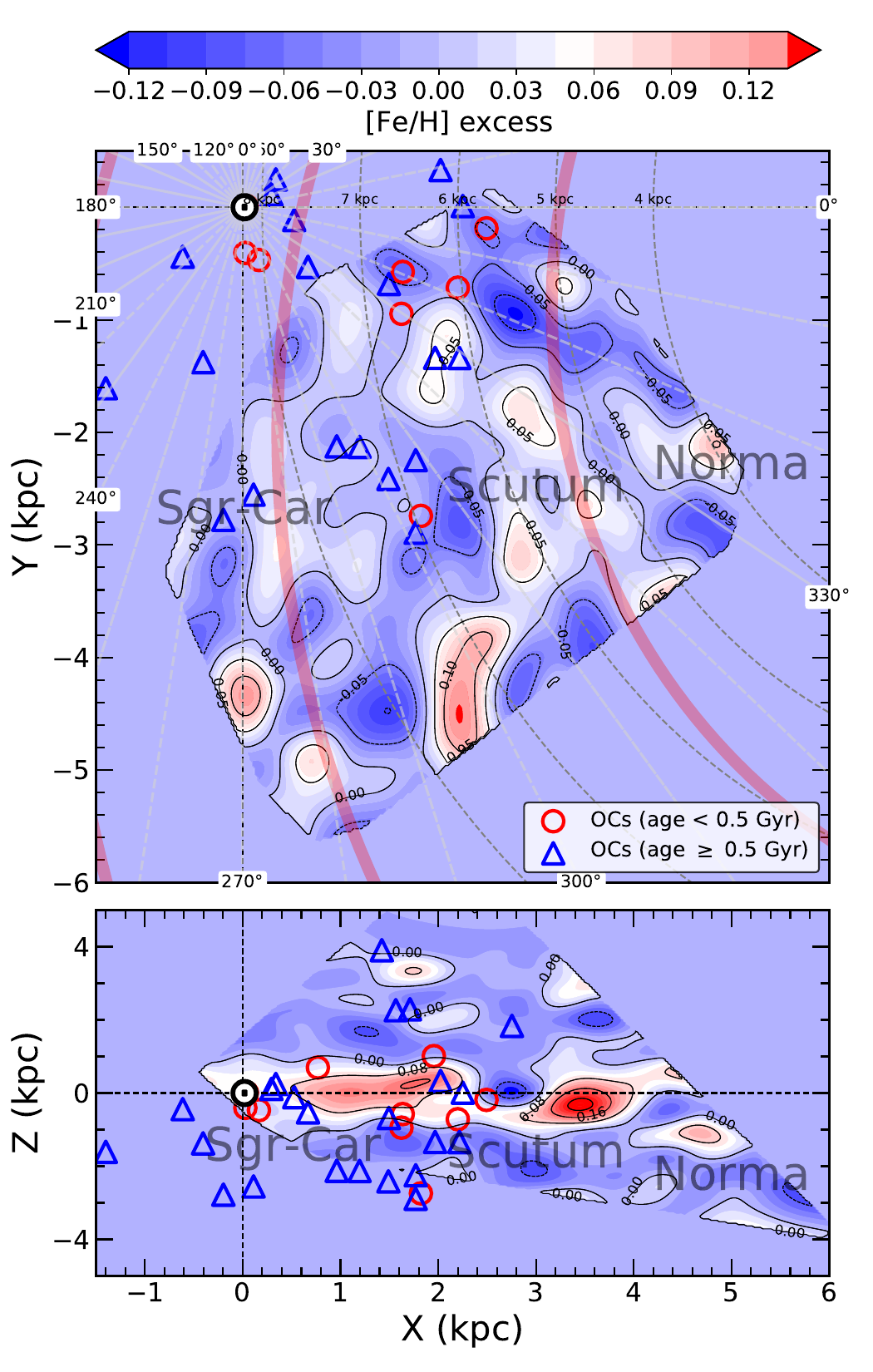}} \caption{Gaussian smoothing of the MSTO GES sample in Cartesian XY and XZ planes, colour-coded by metallicity excess. The overlaid spiral arm model follows the framework provided by \citet{Drimmel2024}. Spatial distribution of star clusters is also shown, colour-coded by age: young clusters (< 0.5 Gyr) as red circles and older clusters ($\geq$ 0.5 Gyr) as blue triangles, plotted in the XY and XZ planes.} \label{fig:XYZ_FEH_excess_clusters} \end{figure}

\begin{figure} \resizebox{\hsize}{!}{\includegraphics{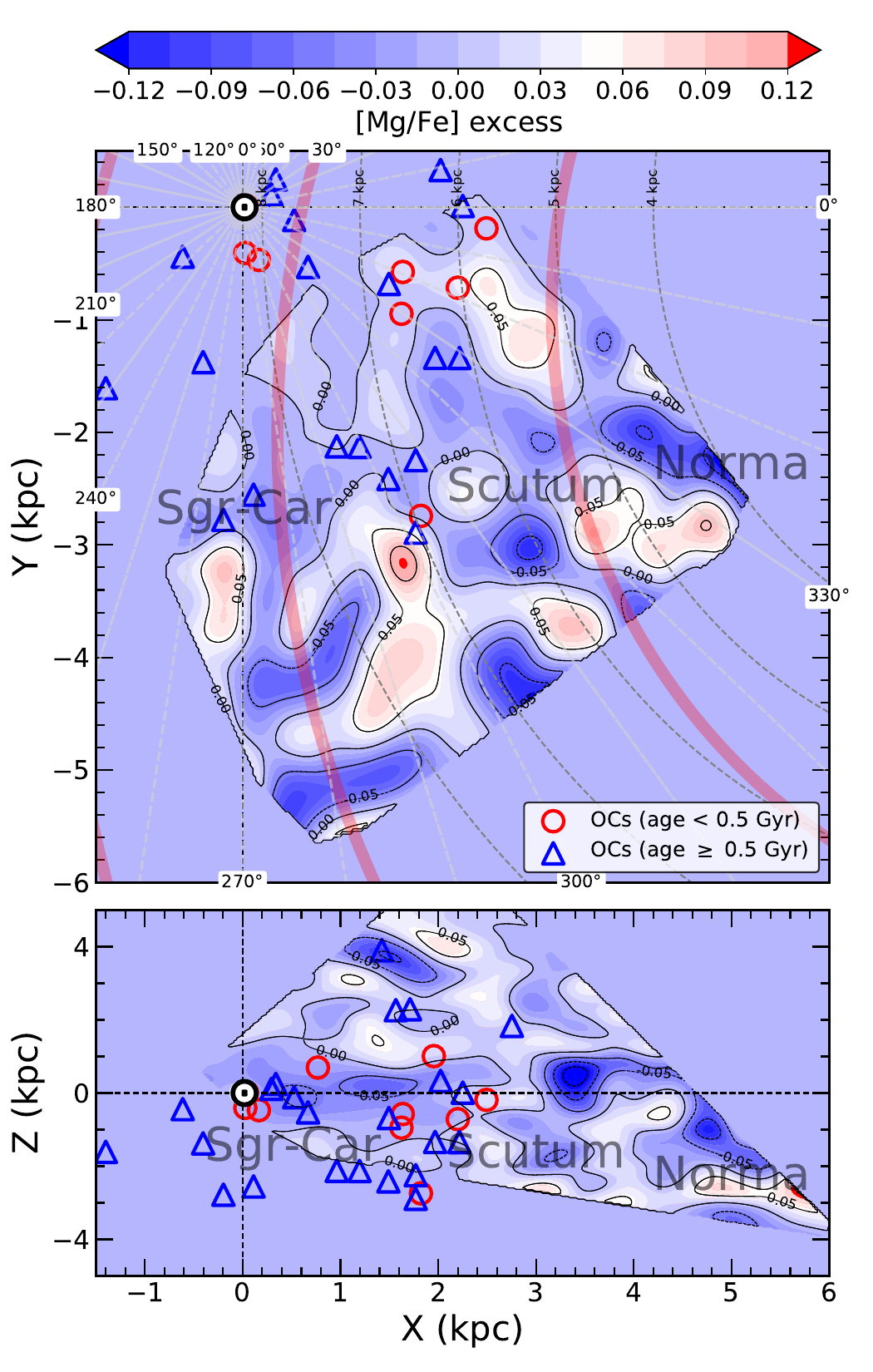}} \caption{Gaussian smoothing of the MSTO GES sample in Cartesian XY and XZ planes, colour-coded by [Mg/Fe] excess. The overlaid spiral arm model follows the framework provided by \citet{Drimmel2024}. Black dots represent the thin disc subsample with ages less than 5 Gyrs.} \label{fig:XYZ_MGFE_excess} \end{figure}

\begin{figure} \resizebox{\hsize}{!}{\includegraphics{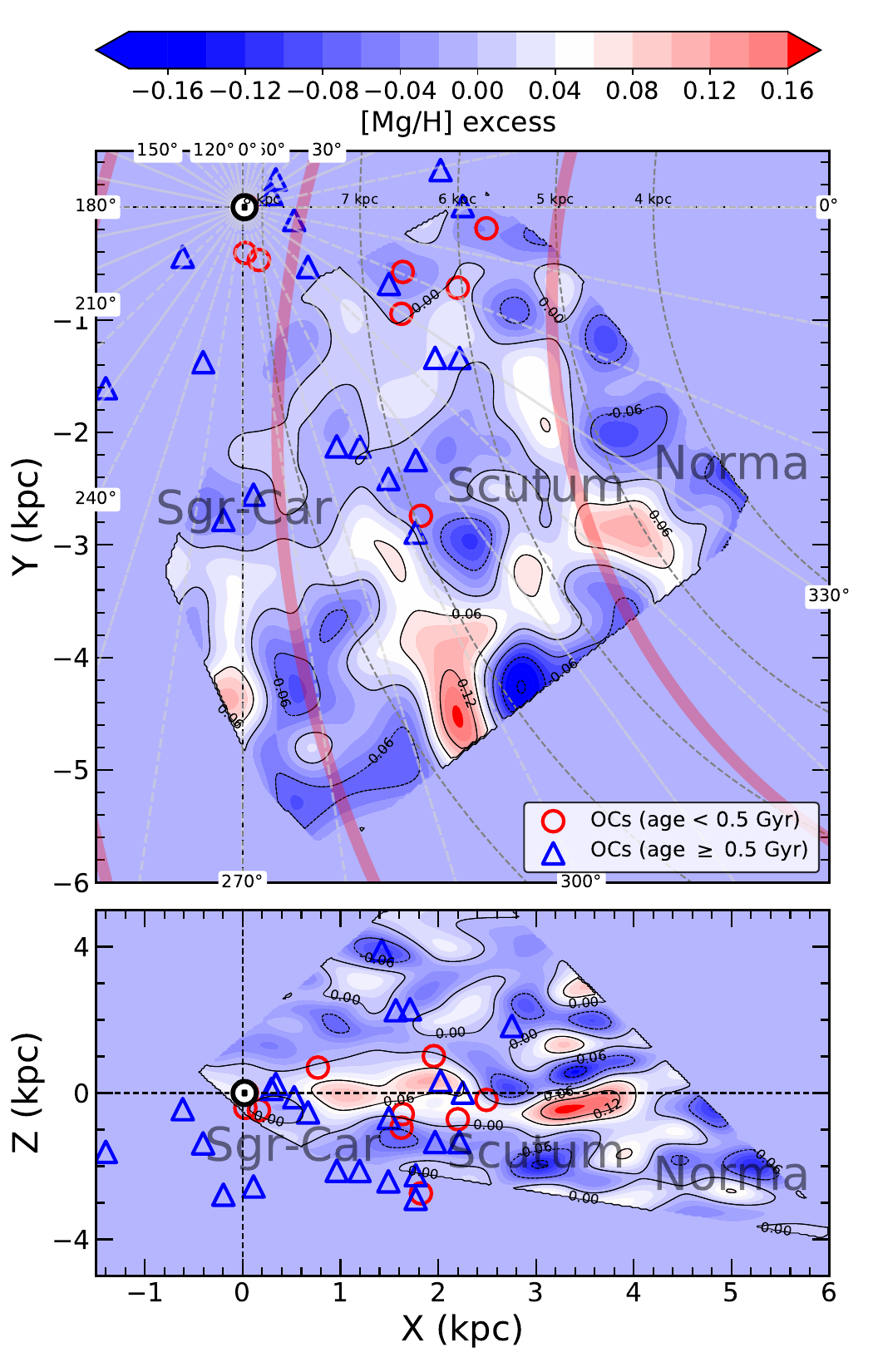}} \caption{Gaussian smoothing of the MSTO GES sample in Cartesian XY and XZ planes, colour-coded by [Mg/H] excess. The overlaid spiral arm model follows the framework provided by \citet{Drimmel2024}. Spatial distribution of star clusters is also shown, colour-coded by age: young clusters (< 0.5 Gyr) as red circles and older clusters ($\geq$ 0.5 Gyr) as blue triangles, plotted in the XY and XZ planes.} \label{fig:XYZ_MGH_excess_clusters} \end{figure}

In addition to the field stars, we have included the GES sample of OCs, which represent a characteristic population of the thin disc \citep[see, e.g.,][]{VanderSwaelmen2024}. These clusters generally have relatively short lifespans and typically dissolve before reaching a few Gyr \citep[see, e.g.,][]{Viscasillas2023, Viscasillas2024}. Figures \ref{fig:XYZ_FEH_excess_clusters}, \ref{fig:XYZ_MGFE_excess}, and \ref{fig:XYZ_MGH_excess_clusters} illustrate that mature OCs (age > 0.5 Gyr) tend to be situated in the inter-arm regions, away from what seems to be the central core of the spiral arms. This pattern is especially evident in the vertical sections of the figures.

\begin{figure}
\centering
   \includegraphics[width=1\hsize]{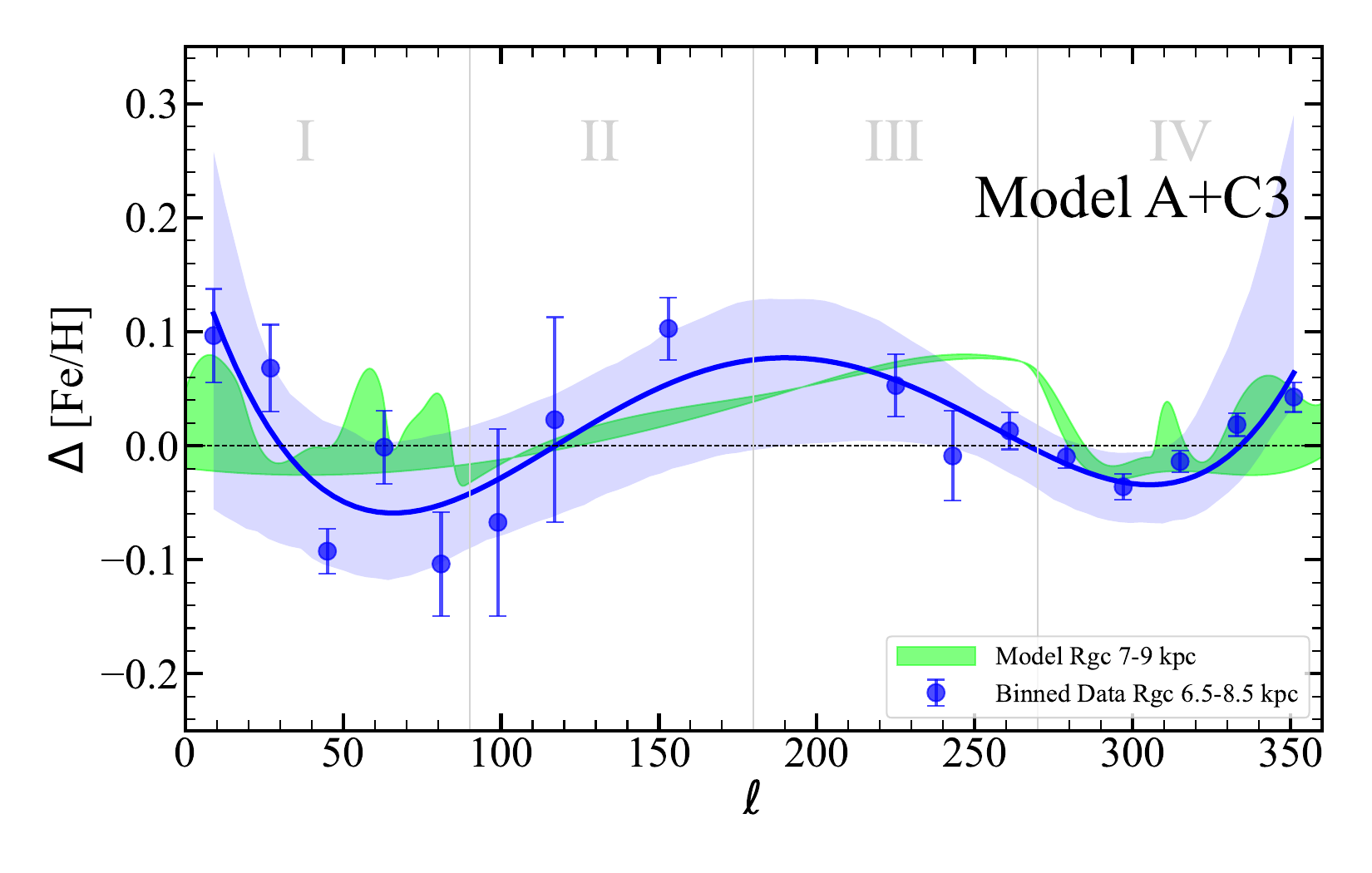}
   \includegraphics[width=1\hsize]{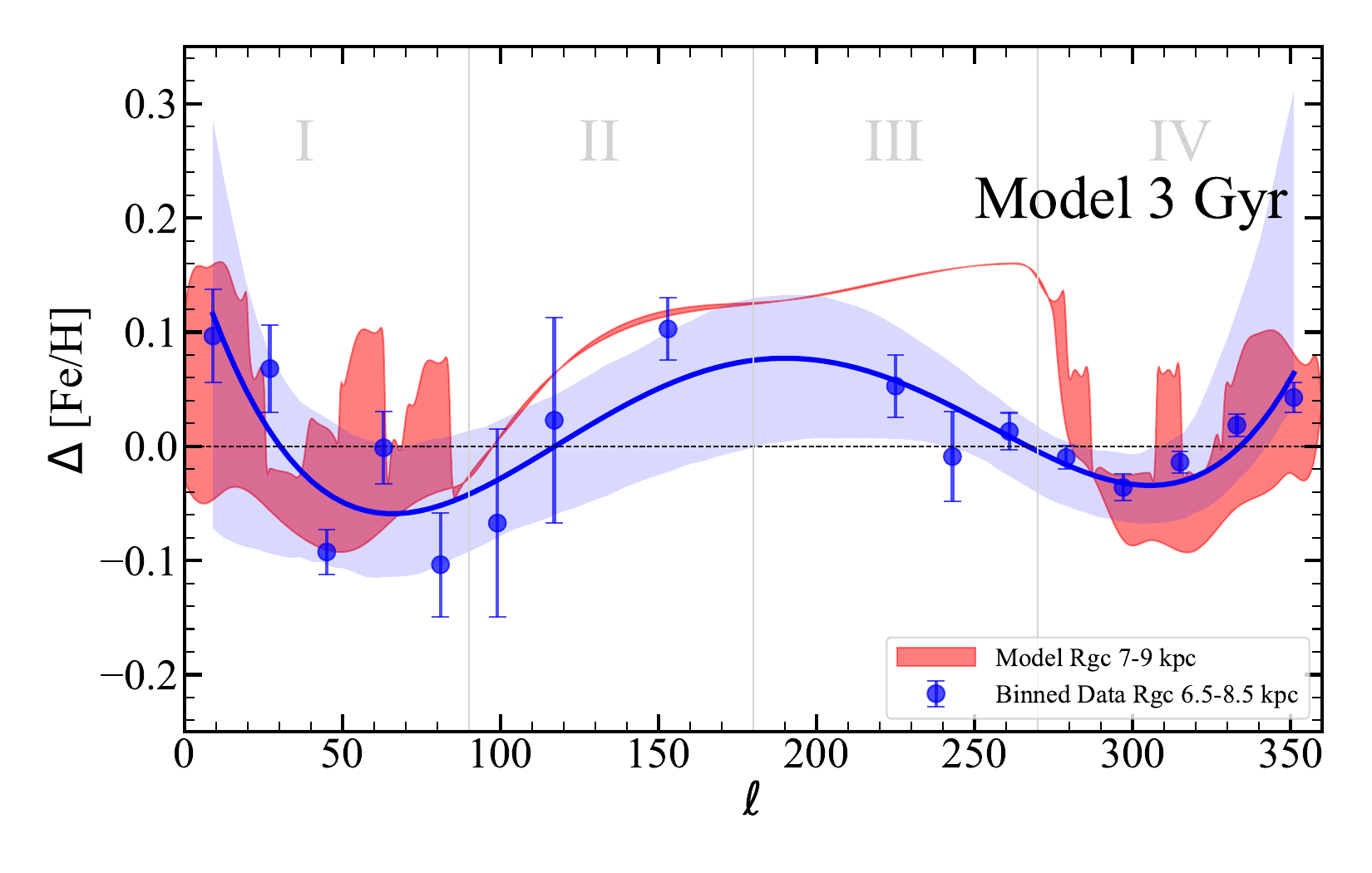}
    \caption{Residuals of [Fe/H] as a function of Galactic longitude ($\ell$) for different radial distances in the Milky Way. The plot shows the binned data (blue scatter) fitted with a polynomial (blue line) with a 95\% confidence interval. Upper panel: data are compared with the Model A+C3 predictions from \citet{Spitoni2023}, which assume that, during the last 1 Gyr of disc evolution, all Galactocentric distances are co-rotations. The model is represented by a dashed green area, indicating the range of residual variations computed between 7 and 9 kpc. The residuals are calculated relative to the mean [Fe/H] value in each considered annular regions. Lower panel: as upper panel but  showing the Model 3 Gyr predictions from \citet{Barbillon2024}, where it is assumed that all Galactocentric distances are co-rotations during the last 3 Gyr of  evolution. The model is represented by a shaded red area.
    } \label{fig:5599_XH_residual_FEH_vs_ModelAC3}
\end{figure}

\begin{figure}
\centering
    \includegraphics[width=1\hsize]{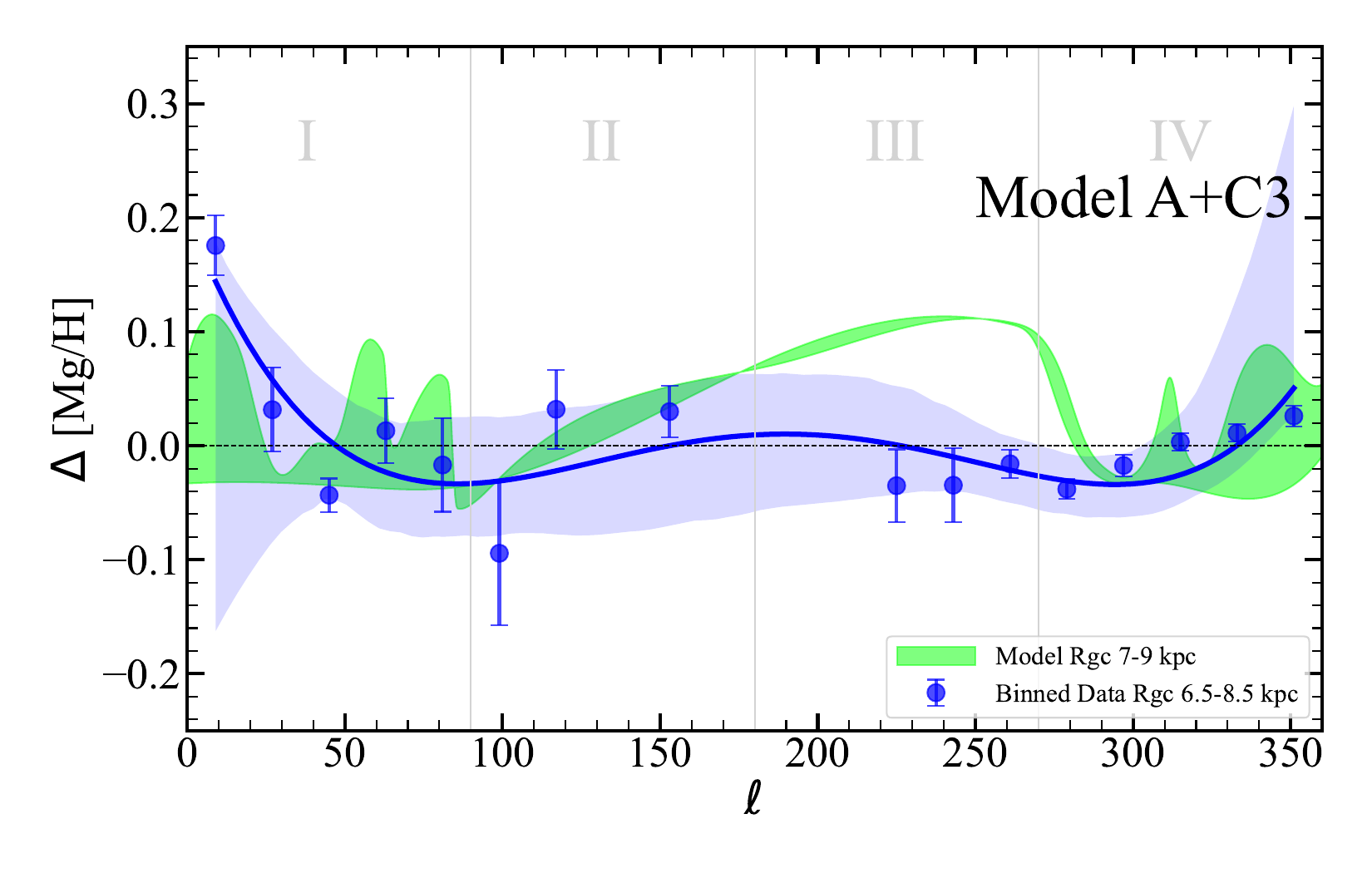} 
     \includegraphics[width=1\hsize]{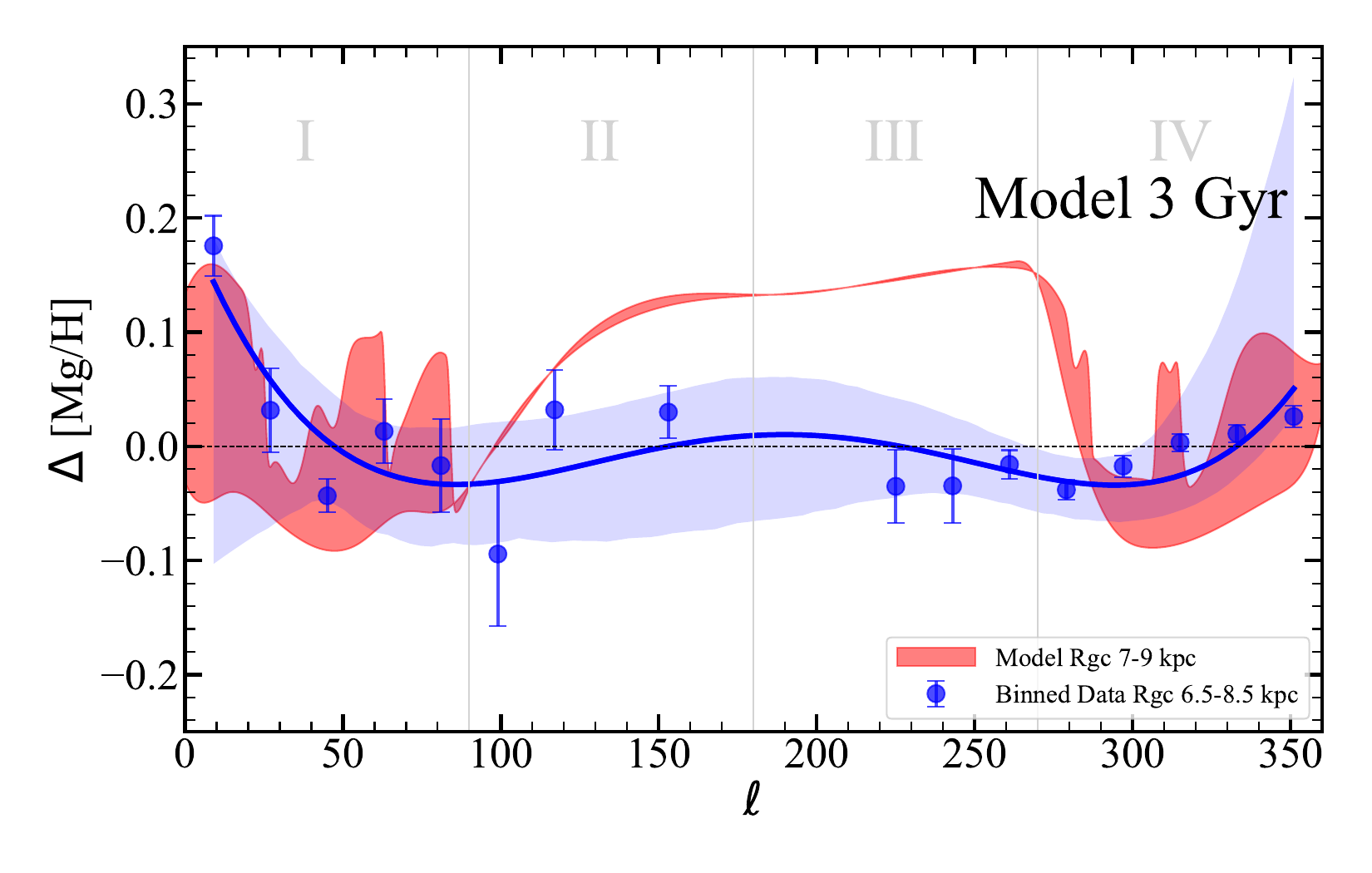}
    \caption{Same as Figure~\ref{fig:5599_XH_residual_FEH_vs_ModelAC3}, but for [Mg/H].} 
    \label{fig:5599_XH_residual_MG1H_vs_ModelAC3} 
\end{figure}

\section{Residual Abundance Variations Across Galactic Longitude: comparison with Chemical Evolution Models}
\label{sec:residuals}

In the literature, several studies have modelled spiral arms as a superposition of multiple spiral density waves with varying pattern speeds \citep[e.g.][]{Tagger1987,Minchev2012}. \citet{Spitoni2023} considered in the 2D chemical evolution model for the Galactic disc introduced by \citet{Spitoni2019}   the spiral structure of the Milky Way as a combination of three distinct segments, each with different pattern speeds, to investigate their effects on the distribution of chemical elements in the Galactic disc. In their model, star formation is enhanced by the passage of spiral arms, leading to abundance fluctuations whose amplitudes depend on the specific chemical element considered. 
The most important result is that elements synthesized on short timescales (i.e., by short-lived progenitors) exhibit larger abundance variations, as they quickly reflect the influence of the spiral arm passage. In contrast, elements released into the interstellar medium after significant delays display smaller abundance variations.
More recently, in \citet{Vasini2025} this model has been adopted to trace the 2D distribution of the short-lived
radioisotope $^{26}$Al.  
In the spiral structures modelled in the 2D chemical evolution studies by \citet{Spitoni2023} and \citet{Barbillon2024}, the authors also considered the possibility of co-rotating arms existing at all radii within transient spiral arm structures, as confirmed by various dynamical studies \citep{Grand2012, Hunt2019}. Supporting these findings, the recent analysis of star clusters by \citet{Liu2025} indicates that the spiral arm pattern speeds align with the Galactic rotation curve.

In particular, in this work, we present results in which for the last 1 Gyr (Model A+C3 of \citealt{Spitoni2023})  and 3 Gyr (Model 3 Gyr of  \citealt{Barbillon2024}) we imposed the condition that all Galactocentric distances serve as loci of co-rotation. This approach is intended to mimic the potential transient nature of this structure.  For a detailed description of the implementation of Galactic spiral arms in a 2D chemical evolution model and the adopted nucleosynthesis prescriptions, we refer the reader to \citet{Spitoni2023}.

Figures \ref{fig:5599_XH_residual_FEH_vs_ModelAC3} and \ref{fig:5599_XH_residual_MG1H_vs_ModelAC3} show the residuals of iron and magnesium abundances, $\Delta$[Fe/H] and $\Delta$[Mg/H], as a function of Galactic longitude ($\ell$). Residuals were computed by subtracting the mean abundance for each azimuthal bin for the MSTO sample of $\sim$5500 stars. The median $R_{\rm GC}$ value in our sample is 7.57 kpc with the first and third quartiles between 6.59-8.46 kpc. The dots represent binned residuals for the most sampled region of our data ($\sim$6.5-8.5 kpc) with a 5th order polynomial fit with a 95\% confidence interval displayed around the regression line, while the overlaid lines correspond to chemical evolution models from \citet[][upper panels]{Spitoni2023} and  \citet[][lower panels]{Barbillon2024} for different Galactocentric distances (between 7 and 9 kpc). Both plots reveal periodic variations in the residuals, which suggest the presence of underlying spiral structures influencing the chemical distribution in the Galactic disc. There is a notable gap in our data sample spanning Galactic longitudes between approximately $150^\circ$ and $210^\circ$. This gap is due to the observational strategy of GES \citep{Gilmore2022}. Fluctuations are observed across Galactic longitude, with a pronounced increase in residuals beyond $300^\circ$. This trend appears to correspond to the regions of the inner Scutum and Norma Arms. At lower Galactic longitudes ($\ell$ < $\sim50^\circ$), the pattern suggests a peak in the residuals, which may be associated with the near side of the Sagittarius Arm.

Among the presented results, Model A+C3 appears to best reproduce the observed abundance variation across the full range of Galactic longitudes. In both models, the variations are more pronounced in quadrant {\sc I} than in quadrant {\sc IV}, aligning with the observational data. However, it is important to note that Model 3 Gyr more accurately accounts for the variations in the observed residuals of both $\Delta$[Mg/H] and $\Delta$[Fe/H] in quadrants {\sc I} and {\sc IV}. However, it is worth noting that several wiggle-like features appear in both Model A+C3 and Model 3 Gyr results. These arise from residual abundance variations computed at different Galactocentric distances across all azimuthal coordinates. Consequently, their projection onto Galactic longitudes leads to the emergence of these patterns. We indeed included all azimuthal coordinates to illustrate the full range of abundance variations induced by the structure of the considered spiral arms. }

The abundance variations revealed by {\it Gaia-ESO} data appear to support the hypothesis of the transient nature of spiral arms. Slight differences emerge in comparison with \citet{Barbillon2024} analysis. In our study, Model 3 Gyr already represents an extreme case, predicting excessively large variations in quadrants {\sc II} and {\sc III}, whereas in \citet{Barbillon2024}, 3 Gyr was the minimum value required to reproduce the deficiency in [Ca/Fe] observed in stars within the spiral arms by Gaia GSP-Spec data. This difference may arise from the distinct nucleosynthesis pathways of Mg and Ca, as they do not share the same proportion of production from Type II and Type Ia supernovae.

\section{Conclusions}
\label{sec:conclusions}
In this paper, we presented the analysis of the azimuthal distributions of [Fe/H], [Mg/H] and [Mg/Fe] in the Galactic disc, using data from the {\it Gaia}-ESO survey. 
Such data extend into the innermost
regions of the disc, complementing previous works \citep[see. e.g.][]{Poggio2021, Poggio2022, Barbillon2024} and confirming that in correspondence with the spiral arms there is an excess of [Fe/H] and underabundance of [Mg/Fe]. Furthermore, we were able to analyse the structure of the arms in the vertical plane, finding a clear decrease in metallicity when moving away from the plane. Our [Mg/H] map  reveals a spur connecting the Sagittarium and Scutum arms.
Finally, we compared the residual structure in [Fe/H] and [Mg/H] as a function of Galactic longitude with two 2D chemical evolution models \citep{Spitoni2023, Barbillon2024}. These models incorporate the effect of spiral arms and assume that all radii are corotating. Our analysis revealed a strong agreement between our observational results and the model predictions, reinforcing the role of spiral arm transits in driving both azimuthal and radial variations in chemical abundances. %

\begin{acknowledgements}
We thank the anonymous referee for her/his useful comments. L.M., C.V.V.,  E.S., G.C., S.R. and G.G.S. thank INAF for the support (Large Grant EPOCH), the Mini-Grants Checs  (1.05.23.04.02), and the financial support under the National Recovery and Resilience Plan (NRRP), Mission 4, Component 2, Investment 1.1, Call for tender No. 104 published on 2.2.2022 by the Italian Ministry of University and Research (MUR), funded by the European Union – NextGenerationEU – Project ‘Cosmic POT’ Grant Assignment Decree No. 2022X4TM3H by the Italian Ministry of the University and Research (MUR). 
A.V. acknowledges financial support from the European Research Council under the ERC Starting Grant "GalFlow" (grant 101116226).
 This work made use of SciPy \citep{Virtanen2020}, Astropy \citep{Astropy_2018}, Seaborn \citep{Waskom2021}, TopCat \citep{Taylor2005}, Pandas \citep{pandas2020} and Matplotlib \citep{Hunter2007}. Based on data products from observations made with ESO
Telescopes at the La Silla Paranal Observatory under programmes 188.B-3002,
193.B-0936, and 197.B-1074.

\end{acknowledgements}

\bibliographystyle{aa} 
\bibliography{Bibliography}

\begin{thebibliography}{59}
\expandafter\ifx\csname natexlab\endcsname\relax\def\natexlab#1{#1}\fi

\bibitem[{{Alinder} {et~al.}(2024){Alinder}, {McMillan}, \& {Bensby}}]{Alinder2024}
{Alinder}, S., {McMillan}, P.~J., \& {Bensby}, T. 2024, \aap, 690, A15

\bibitem[{{Astropy Collaboration} {et~al.}(2018){Astropy Collaboration}, {Price-Whelan}, {Sip{\H{o}}cz}, {G{\"u}nther}, {Lim}, {Crawford}, {Conseil}, {Shupe}, {Craig}, {Dencheva}, {Ginsburg}, {VanderPlas}, {Bradley}, {P{\'e}rez-Su{\'a}rez}, {de Val-Borro}, {Aldcroft}, {Cruz}, {Robitaille}, {Tollerud}, {Ardelean}, {Babej}, {Bach}, {Bachetti}, {Bakanov}, {Bamford}, {Barentsen}, {Barmby}, {Baumbach}, {Berry}, {Biscani}, {Boquien}, {Bostroem}, {Bouma}, {Brammer}, {Bray}, {Breytenbach}, {Buddelmeijer}, {Burke}, {Calderone}, {Cano Rodr{\'\i}guez}, {Cara}, {Cardoso}, {Cheedella}, {Copin}, {Corrales}, {Crichton}, {D'Avella}, {Deil}, {Depagne}, {Dietrich}, {Donath}, {Droettboom}, {Earl}, {Erben}, {Fabbro}, {Ferreira}, {Finethy}, {Fox}, {Garrison}, {Gibbons}, {Goldstein}, {Gommers}, {Greco}, {Greenfield}, {Groener}, {Grollier}, {Hagen}, {Hirst}, {Homeier}, {Horton}, {Hosseinzadeh}, {Hu}, {Hunkeler}, {Ivezi{\'c}}, {Jain}, {Jenness}, {Kanarek}, {Kendrew}, {Kern}, {Kerzendorf}, {Khvalko}, {King}, {Kirkby}, {Kulkarni},
  {Kumar}, {Lee}, {Lenz}, {Littlefair}, {Ma}, {Macleod}, {Mastropietro}, {McCully}, {Montagnac}, {Morris}, {Mueller}, {Mumford}, {Muna}, {Murphy}, {Nelson}, {Nguyen}, {Ninan}, {N{\"o}the}, {Ogaz}, {Oh}, {Parejko}, {Parley}, {Pascual}, {Patil}, {Patil}, {Plunkett}, {Prochaska}, {Rastogi}, {Reddy Janga}, {Sabater}, {Sakurikar}, {Seifert}, {Sherbert}, {Sherwood-Taylor}, {Shih}, {Sick}, {Silbiger}, {Singanamalla}, {Singer}, {Sladen}, {Sooley}, {Sornarajah}, {Streicher}, {Teuben}, {Thomas}, {Tremblay}, {Turner}, {Terr{\'o}n}, {van Kerkwijk}, {de la Vega}, {Watkins}, {Weaver}, {Whitmore}, {Woillez}, {Zabalza}, \& {Astropy Contributors}}]{Astropy_2018}
{Astropy Collaboration}, {Price-Whelan}, A.~M., {Sip{\H{o}}cz}, B.~M., {et~al.} 2018, \aj, 156, 123

\bibitem[{{Barbillon} {et~al.}(2025){Barbillon}, {Recio-Blanco}, {Poggio}, {Palicio}, {Spitoni}, {de Laverny}, \& {Cescutti}}]{Barbillon2024}
{Barbillon}, M., {Recio-Blanco}, A., {Poggio}, E., {et~al.} 2025, \aap, 693, A3

\bibitem[{{Bovy}(2015)}]{bovy15}
{Bovy}, J. 2015, \apjs, 216, 29

\bibitem[{{Bressan} {et~al.}(2012){Bressan}, {Marigo}, {Girardi}, {Salasnich}, {Dal Cero}, {Rubele}, \& {Nanni}}]{Bressan2012}
{Bressan}, A., {Marigo}, P., {Girardi}, L., {et~al.} 2012, \mnras, 427, 127

\bibitem[{{Candebat} {et~al.}(2024){Candebat}, {Germano Sacco}, {Magrini}, {Belfiore}, {Van-der-Swaelmen}, \& {Zibetti}}]{Candebat2024}
{Candebat}, N., {Germano Sacco}, G., {Magrini}, L., {et~al.} 2024, arXiv e-prints, arXiv:2409.10621

\bibitem[{{Castro-Ginard} {et~al.}(2021){Castro-Ginard}, {McMillan}, {Luri}, {Jordi}, {Romero-G{\'o}mez}, {Cantat-Gaudin}, {Casamiquela}, {Tarricq}, {Soubiran}, \& {Anders}}]{Castro-Ginard2021}
{Castro-Ginard}, A., {McMillan}, P.~J., {Luri}, X., {et~al.} 2021, \aap, 652, A162

\bibitem[{{Chieffi} \& {Limongi}(2004)}]{Chieffi2004ApJ...608..405C}
{Chieffi}, A. \& {Limongi}, M. 2004, \apj, 608, 405

\bibitem[{{Chieffi} \& {Limongi}(2013)}]{Chieffi2013ApJ...764...21C}
{Chieffi}, A. \& {Limongi}, M. 2013, \apj, 764, 21

\bibitem[{{Dobbs} \& {Bonnell}(2006)}]{Dobbs2006}
{Dobbs}, C.~L. \& {Bonnell}, I.~A. 2006, \mnras, 367, 873

\bibitem[{{Drimmel} {et~al.}(2024){Drimmel}, {Khanna}, {Poggio}, \& {Skowron}}]{Drimmel2024}
{Drimmel}, R., {Khanna}, S., {Poggio}, E., \& {Skowron}, D.~M. 2024, arXiv e-prints, arXiv:2406.09127

\bibitem[{{Elmegreen}(1980)}]{Elmegreen1980}
{Elmegreen}, D.~M. 1980, \apj, 242, 528

\bibitem[{{Gaia Collaboration} {et~al.}(2021){Gaia Collaboration}, {Brown}, {Vallenari}, {Prusti}, {de Bruijne}, {Babusiaux}, {Biermann}, {Creevey}, {Evans}, {Eyer}, {Hutton}, {Jansen}, {Jordi}, {Klioner}, {Lammers}, {Lindegren}, {Luri}, {Mignard}, {Panem}, {Pourbaix}, {Randich}, {Sartoretti}, {Soubiran}, {Walton}, {Arenou}, {Bailer-Jones}, {Bastian}, {Cropper}, {Drimmel}, {Katz}, {Lattanzi}, {van Leeuwen}, {Bakker}, {Cacciari}, {Casta{\~n}eda}, {De Angeli}, {Ducourant}, {Fabricius}, {Fouesneau}, {Fr{\'e}mat}, {Guerra}, {Guerrier}, {Guiraud}, {Jean-Antoine Piccolo}, {Masana}, {Messineo}, {Mowlavi}, {Nicolas}, {Nienartowicz}, {Pailler}, {Panuzzo}, {Riclet}, {Roux}, {Seabroke}, {Sordo}, {Tanga}, {Th{\'e}venin}, {Gracia-Abril}, {Portell}, {Teyssier}, {Altmann}, {Andrae}, {Bellas-Velidis}, {Benson}, {Berthier}, {Blomme}, {Brugaletta}, {Burgess}, {Busso}, {Carry}, {Cellino}, {Cheek}, {Clementini}, {Damerdji}, {Davidson}, {Delchambre}, {Dell'Oro}, {Fern{\'a}ndez-Hern{\'a}ndez}, {Galluccio}, {Garc{\'\i}a-Lario},
  {Garcia-Reinaldos}, {Gonz{\'a}lez-N{\'u}{\~n}ez}, {Gosset}, {Haigron}, {Halbwachs}, {Hambly}, {Harrison}, {Hatzidimitriou}, {Heiter}, {Hern{\'a}ndez}, {Hestroffer}, {Hodgkin}, {Holl}, {Jan{\ss}en}, {Jevardat de Fombelle}, {Jordan}, {Krone-Martins}, {Lanzafame}, {L{\"o}ffler}, {Lorca}, {Manteiga}, {Marchal}, {Marrese}, {Moitinho}, {Mora}, {Muinonen}, {Osborne}, {Pancino}, {Pauwels}, {Petit}, {Recio-Blanco}, {Richards}, {Riello}, {Rimoldini}, {Robin}, {Roegiers}, {Rybizki}, {Sarro}, {Siopis}, {Smith}, {Sozzetti}, {Ulla}, {Utrilla}, {van Leeuwen}, {van Reeven}, {Abbas}, {Abreu Aramburu}, {Accart}, {Aerts}, {Aguado}, {Ajaj}, {Altavilla}, {{\'A}lvarez}, {{\'A}lvarez Cid-Fuentes}, {Alves}, {Anderson}, {Anglada Varela}, {Antoja}, {Audard}, {Baines}, {Baker}, {Balaguer-N{\'u}{\~n}ez}, {Balbinot}, {Balog}, {Barache}, {Barbato}, {Barros}, {Barstow}, {Bartolom{\'e}}, {Bassilana}, {Bauchet}, {Baudesson-Stella}, {Becciani}, {Bellazzini}, {Bernet}, {Bertone}, {Bianchi}, {Blanco-Cuaresma}, {Boch}, {Bombrun}, {Bossini},
  {Bouquillon}, {Bragaglia}, {Bramante}, {Breedt}, {Bressan}, {Brouillet}, {Bucciarelli}, {Burlacu}, {Busonero}, {Butkevich}, {Buzzi}, {Caffau}, {Cancelliere}, {C{\'a}novas}, {Cantat-Gaudin}, {Carballo}, {Carlucci}, {Carnerero}, {Carrasco}, {Casamiquela}, {Castellani}, {Castro-Ginard}, {Castro Sampol}, {Chaoul}, {Charlot}, {Chemin}, {Chiavassa}, {Cioni}, {Comoretto}, {Cooper}, {Cornez}, {Cowell}, {Crifo}, {Crosta}, {Crowley}, {Dafonte}, {Dapergolas}, {David}, {David}, {de Laverny}, {De Luise}, {De March}, {De Ridder}, {de Souza}, {de Teodoro}, {de Torres}, {del Peloso}, {del Pozo}, {Delbo}, {Delgado}, {Delgado}, {Delisle}, {Di Matteo}, {Diakite}, {Diener}, {Distefano}, {Dolding}, {Eappachen}, {Edvardsson}, {Enke}, {Esquej}, {Fabre}, {Fabrizio}, {Faigler}, {Fedorets}, {Fernique}, {Fienga}, {Figueras}, {Fouron}, {Fragkoudi}, {Fraile}, {Franke}, {Gai}, {Garabato}, {Garcia-Gutierrez}, {Garc{\'\i}a-Torres}, {Garofalo}, {Gavras}, {Gerlach}, {Geyer}, {Giacobbe}, {Gilmore}, {Girona}, {Giuffrida}, {Gomel}, {Gomez},
  {Gonzalez-Santamaria}, {Gonz{\'a}lez-Vidal}, {Granvik}, {Guti{\'e}rrez-S{\'a}nchez}, {Guy}, {Hauser}, {Haywood}, {Helmi}, {Hidalgo}, {Hilger}, {H{\l}adczuk}, {Hobbs}, {Holland}, {Huckle}, {Jasniewicz}, {Jonker}, {Juaristi Campillo}, {Julbe}, {Karbevska}, {Kervella}, {Khanna}, {Kochoska}, {Kontizas}, {Kordopatis}, {Korn}, {Kostrzewa-Rutkowska}, {Kruszy{\'n}ska}, {Lambert}, {Lanza}, {Lasne}, {Le Campion}, {Le Fustec}, {Lebreton}, {Lebzelter}, {Leccia}, {Leclerc}, {Lecoeur-Taibi}, {Liao}, {Licata}, {Lindstr{\o}m}, {Lister}, {Livanou}, {Lobel}, {Madrero Pardo}, {Managau}, {Mann}, {Marchant}, {Marconi}, {Marcos Santos}, {Marinoni}, {Marocco}, {Marshall}, {Martin Polo}, {Mart{\'\i}n-Fleitas}, {Masip}, {Massari}, {Mastrobuono-Battisti}, {Mazeh}, {McMillan}, {Messina}, {Michalik}, {Millar}, {Mints}, {Molina}, {Molinaro}, {Moln{\'a}r}, {Montegriffo}, {Mor}, {Morbidelli}, {Morel}, {Morris}, {Mulone}, {Munoz}, {Muraveva}, {Murphy}, {Musella}, {Noval}, {Ord{\'e}novic}, {Orr{\`u}}, {Osinde}, {Pagani}, {Pagano},
  {Palaversa}, {Palicio}, {Panahi}, {Pawlak}, {Pe{\~n}alosa Esteller}, {Penttil{\"a}}, {Piersimoni}, {Pineau}, {Plachy}, {Plum}, {Poggio}, {Poretti}, {Poujoulet}, {Pr{\v{s}}a}, {Pulone}, {Racero}, {Ragaini}, {Rainer}, {Raiteri}, {Rambaux}, {Ramos}, {Ramos-Lerate}, {Re Fiorentin}, {Regibo}, {Reyl{\'e}}, {Ripepi}, {Riva}, {Rixon}, {Robichon}, {Robin}, {Roelens}, {Rohrbasser}, {Romero-G{\'o}mez}, {Rowell}, {Royer}, {Rybicki}, {Sadowski}, {Sagrist{\`a} Sell{\'e}s}, {Sahlmann}, {Salgado}, {Salguero}, {Samaras}, {Sanchez Gimenez}, {Sanna}, {Santove{\~n}a}, {Sarasso}, {Schultheis}, {Sciacca}, {Segol}, {Segovia}, {S{\'e}gransan}, {Semeux}, {Shahaf}, {Siddiqui}, {Siebert}, {Siltala}, {Slezak}, {Smart}, {Solano}, {Solitro}, {Souami}, {Souchay}, {Spagna}, {Spoto}, {Steele}, {Steidelm{\"u}ller}, {Stephenson}, {S{\"u}veges}, {Szabados}, {Szegedi-Elek}, {Taris}, {Tauran}, {Taylor}, {Teixeira}, {Thuillot}, {Tonello}, {Torra}, {Torra}, {Turon}, {Unger}, {Vaillant}, {van Dillen}, {Vanel}, {Vecchiato}, {Viala}, {Vicente},
  {Voutsinas}, {Weiler}, {Wevers}, {Wyrzykowski}, {Yoldas}, {Yvard}, {Zhao}, {Zorec}, {Zucker}, {Zurbach}, \& {Zwitter}}]{GaiaCollaboration2021}
{Gaia Collaboration}, {Brown}, A.~G.~A., {Vallenari}, A., {et~al.} 2021, \aap, 649, A1

\bibitem[{{Gaia Collaboration} {et~al.}(2023){Gaia Collaboration}, {Vallenari}, {Brown}, {Prusti}, {de Bruijne}, {Arenou}, {Babusiaux}, {Biermann}, {Creevey}, {Ducourant}, {Evans}, {Eyer}, {Guerra}, {Hutton}, {Jordi}, {Klioner}, {Lammers}, {Lindegren}, {Luri}, {Mignard}, {Panem}, {Pourbaix}, {Randich}, {Sartoretti}, {Soubiran}, {Tanga}, {Walton}, {Bailer-Jones}, {Bastian}, {Drimmel}, {Jansen}, {Katz}, {Lattanzi}, {van Leeuwen}, {Bakker}, {Cacciari}, {Casta{\~n}eda}, {De Angeli}, {Fabricius}, {Fouesneau}, {Fr{\'e}mat}, {Galluccio}, {Guerrier}, {Heiter}, {Masana}, {Messineo}, {Mowlavi}, {Nicolas}, {Nienartowicz}, {Pailler}, {Panuzzo}, {Riclet}, {Roux}, {Seabroke}, {Sordo}, {Th{\'e}venin}, {Gracia-Abril}, {Portell}, {Teyssier}, {Altmann}, {Andrae}, {Audard}, {Bellas-Velidis}, {Benson}, {Berthier}, {Blomme}, {Burgess}, {Busonero}, {Busso}, {C{\'a}novas}, {Carry}, {Cellino}, {Cheek}, {Clementini}, {Damerdji}, {Davidson}, {de Teodoro}, {Nu{\~n}ez Campos}, {Delchambre}, {Dell'Oro}, {Esquej},
  {Fern{\'a}ndez-Hern{\'a}ndez}, {Fraile}, {Garabato}, {Garc{\'\i}a-Lario}, {Gosset}, {Haigron}, {Halbwachs}, {Hambly}, {Harrison}, {Hern{\'a}ndez}, {Hestroffer}, {Hodgkin}, {Holl}, {Jan{\ss}en}, {Jevardat de Fombelle}, {Jordan}, {Krone-Martins}, {Lanzafame}, {L{\"o}ffler}, {Marchal}, {Marrese}, {Moitinho}, {Muinonen}, {Osborne}, {Pancino}, {Pauwels}, {Recio-Blanco}, {Reyl{\'e}}, {Riello}, {Rimoldini}, {Roegiers}, {Rybizki}, {Sarro}, {Siopis}, {Smith}, {Sozzetti}, {Utrilla}, {van Leeuwen}, {Abbas}, {{\'A}brah{\'a}m}, {Abreu Aramburu}, {Aerts}, {Aguado}, {Ajaj}, {Aldea-Montero}, {Altavilla}, {{\'A}lvarez}, {Alves}, {Anders}, {Anderson}, {Anglada Varela}, {Antoja}, {Baines}, {Baker}, {Balaguer-N{\'u}{\~n}ez}, {Balbinot}, {Balog}, {Barache}, {Barbato}, {Barros}, {Barstow}, {Bartolom{\'e}}, {Bassilana}, {Bauchet}, {Becciani}, {Bellazzini}, {Berihuete}, {Bernet}, {Bertone}, {Bianchi}, {Binnenfeld}, {Blanco-Cuaresma}, {Blazere}, {Boch}, {Bombrun}, {Bossini}, {Bouquillon}, {Bragaglia}, {Bramante}, {Breedt},
  {Bressan}, {Brouillet}, {Brugaletta}, {Bucciarelli}, {Burlacu}, {Butkevich}, {Buzzi}, {Caffau}, {Cancelliere}, {Cantat-Gaudin}, {Carballo}, {Carlucci}, {Carnerero}, {Carrasco}, {Casamiquela}, {Castellani}, {Castro-Ginard}, {Chaoul}, {Charlot}, {Chemin}, {Chiaramida}, {Chiavassa}, {Chornay}, {Comoretto}, {Contursi}, {Cooper}, {Cornez}, {Cowell}, {Crifo}, {Cropper}, {Crosta}, {Crowley}, {Dafonte}, {Dapergolas}, {David}, {David}, {de Laverny}, {De Luise}, {De March}, {De Ridder}, {de Souza}, {de Torres}, {del Peloso}, {del Pozo}, {Delbo}, {Delgado}, {Delisle}, {Demouchy}, {Dharmawardena}, {Di Matteo}, {Diakite}, {Diener}, {Distefano}, {Dolding}, {Edvardsson}, {Enke}, {Fabre}, {Fabrizio}, {Faigler}, {Fedorets}, {Fernique}, {Fienga}, {Figueras}, {Fournier}, {Fouron}, {Fragkoudi}, {Gai}, {Garcia-Gutierrez}, {Garcia-Reinaldos}, {Garc{\'\i}a-Torres}, {Garofalo}, {Gavel}, {Gavras}, {Gerlach}, {Geyer}, {Giacobbe}, {Gilmore}, {Girona}, {Giuffrida}, {Gomel}, {Gomez}, {Gonz{\'a}lez-N{\'u}{\~n}ez},
  {Gonz{\'a}lez-Santamar{\'\i}a}, {Gonz{\'a}lez-Vidal}, {Granvik}, {Guillout}, {Guiraud}, {Guti{\'e}rrez-S{\'a}nchez}, {Guy}, {Hatzidimitriou}, {Hauser}, {Haywood}, {Helmer}, {Helmi}, {Sarmiento}, {Hidalgo}, {Hilger}, {H{\l}adczuk}, {Hobbs}, {Holland}, {Huckle}, {Jardine}, {Jasniewicz}, {Jean-Antoine Piccolo}, {Jim{\'e}nez-Arranz}, {Jorissen}, {Juaristi Campillo}, {Julbe}, {Karbevska}, {Kervella}, {Khanna}, {Kontizas}, {Kordopatis}, {Korn}, {K{\'o}sp{\'a}l}, {Kostrzewa-Rutkowska}, {Kruszy{\'n}ska}, {Kun}, {Laizeau}, {Lambert}, {Lanza}, {Lasne}, {Le Campion}, {Lebreton}, {Lebzelter}, {Leccia}, {Leclerc}, {Lecoeur-Taibi}, {Liao}, {Licata}, {Lindstr{\o}m}, {Lister}, {Livanou}, {Lobel}, {Lorca}, {Loup}, {Madrero Pardo}, {Magdaleno Romeo}, {Managau}, {Mann}, {Manteiga}, {Marchant}, {Marconi}, {Marcos}, {Marcos Santos}, {Mar{\'\i}n Pina}, {Marinoni}, {Marocco}, {Marshall}, {Martin Polo}, {Mart{\'\i}n-Fleitas}, {Marton}, {Mary}, {Masip}, {Massari}, {Mastrobuono-Battisti}, {Mazeh}, {McMillan}, {Messina}, {Michalik},
  {Millar}, {Mints}, {Molina}, {Molinaro}, {Moln{\'a}r}, {Monari}, {Mongui{\'o}}, {Montegriffo}, {Montero}, {Mor}, {Mora}, {Morbidelli}, {Morel}, {Morris}, {Muraveva}, {Murphy}, {Musella}, {Nagy}, {Noval}, {Oca{\~n}a}, {Ogden}, {Ordenovic}, {Osinde}, {Pagani}, {Pagano}, {Palaversa}, {Palicio}, {Pallas-Quintela}, {Panahi}, {Payne-Wardenaar}, {Pe{\~n}alosa Esteller}, {Penttil{\"a}}, {Pichon}, {Piersimoni}, {Pineau}, {Plachy}, {Plum}, {Poggio}, {Pr{\v{s}}a}, {Pulone}, {Racero}, {Ragaini}, {Rainer}, {Raiteri}, {Rambaux}, {Ramos}, {Ramos-Lerate}, {Re Fiorentin}, {Regibo}, {Richards}, {Rios Diaz}, {Ripepi}, {Riva}, {Rix}, {Rixon}, {Robichon}, {Robin}, {Robin}, {Roelens}, {Rogues}, {Rohrbasser}, {Romero-G{\'o}mez}, {Rowell}, {Royer}, {Ruz Mieres}, {Rybicki}, {Sadowski}, {S{\'a}ez N{\'u}{\~n}ez}, {Sagrist{\`a} Sell{\'e}s}, {Sahlmann}, {Salguero}, {Samaras}, {Sanchez Gimenez}, {Sanna}, {Santove{\~n}a}, {Sarasso}, {Schultheis}, {Sciacca}, {Segol}, {Segovia}, {S{\'e}gransan}, {Semeux}, {Shahaf}, {Siddiqui}, {Siebert},
  {Siltala}, {Silvelo}, {Slezak}, {Slezak}, {Smart}, {Snaith}, {Solano}, {Solitro}, {Souami}, {Souchay}, {Spagna}, {Spina}, {Spoto}, {Steele}, {Steidelm{\"u}ller}, {Stephenson}, {S{\"u}veges}, {Surdej}, {Szabados}, {Szegedi-Elek}, {Taris}, {Taylor}, {Teixeira}, {Tolomei}, {Tonello}, {Torra}, {Torra}, {Torralba Elipe}, {Trabucchi}, {Tsounis}, {Turon}, {Ulla}, {Unger}, {Vaillant}, {van Dillen}, {van Reeven}, {Vanel}, {Vecchiato}, {Viala}, {Vicente}, {Voutsinas}, {Weiler}, {Wevers}, {Wyrzykowski}, {Yoldas}, {Yvard}, {Zhao}, {Zorec}, {Zucker}, \& {Zwitter}}]{GaiaCollaboration2023}
{Gaia Collaboration}, {Vallenari}, A., {Brown}, A.~G.~A., {et~al.} 2023, \aap, 674, A1

\bibitem[{{Ge} {et~al.}(2024){Ge}, {Li}, {Hao}, {Lin}, {Hou}, {Liu}, {Li}, \& {Bian}}]{Ge_2024}
{Ge}, Q.~A., {Li}, J.~J., {Hao}, C.~J., {et~al.} 2024, \aj, 168, 25

\bibitem[{{Gilmore} {et~al.}(2022){Gilmore}, {Randich}, {Worley}, {Hourihane}, {Gonneau}, {Sacco}, {Lewis}, {Magrini}, {Fran{\c{c}}ois}, {Jeffries}, {Koposov}, {Bragaglia}, {Alfaro}, {Allende Prieto}, {Blomme}, {Korn}, {Lanzafame}, {Pancino}, {Recio-Blanco}, {Smiljanic}, {Van Eck}, {Zwitter}, {Bensby}, {Flaccomio}, {Irwin}, {Franciosini}, {Morbidelli}, {Damiani}, {Bonito}, {Friel}, {Vink}, {Prisinzano}, {Abbas}, {Hatzidimitriou}, {Held}, {Jordi}, {Paunzen}, {Spagna}, {Jackson}, {Ma{\'\i}z Apell{\'a}niz}, {Asplund}, {Bonifacio}, {Feltzing}, {Binney}, {Drew}, {Ferguson}, {Micela}, {Negueruela}, {Prusti}, {Rix}, {Vallenari}, {Bergemann}, {Casey}, {de Laverny}, {Frasca}, {Hill}, {Lind}, {Sbordone}, {Sousa}, {Adibekyan}, {Caffau}, {Daflon}, {Feuillet}, {Gebran}, {Gonzalez Hernandez}, {Guiglion}, {Herrero}, {Lobel}, {Merle}, {Mikolaitis}, {Montes}, {Morel}, {Ruchti}, {Soubiran}, {Tabernero}, {Tautvai{\v{s}}ien{\.{e}}}, {Traven}, {Valentini}, {Van der Swaelmen}, {Villanova}, {Viscasillas V{\'a}zquez}, {Bayo},
  {Biazzo}, {Carraro}, {Edvardsson}, {Heiter}, {Jofr{\'e}}, {Marconi}, {Martayan}, {Masseron}, {Monaco}, {Walton}, {Zaggia}, {Aguirre B{\o}rsen-Koch}, {Alves}, {Balaguer-Nunez}, {Barklem}, {Barrado}, {Bellazzini}, {Berlanas}, {Binks}, {Bressan}, {Capuzzo-Dolcetta}, {Casagrande}, {Casamiquela}, {Collins}, {D'Orazi}, {Dantas}, {Debattista}, {Delgado-Mena}, {Di Marcantonio}, {Drazdauskas}, {Evans}, {Famaey}, {Franchini}, {Fr{\'e}mat}, {Fu}, {Geisler}, {Gerhard}, {Gonz{\'a}lez Solares}, {Grebel}, {Guti{\'e}rrez Albarr{\'a}n}, {Jim{\'e}nez-Esteban}, {J{\"o}nsson}, {Khachaturyants}, {Kordopatis}, {Kos}, {Lagarde}, {Ludwig}, {Mahy}, {Mapelli}, {Marfil}, {Martell}, {Messina}, {Miglio}, {Minchev}, {Moitinho}, {Montalban}, {Monteiro}, {Morossi}, {Mowlavi}, {Mucciarelli}, {Murphy}, {Nardetto}, {Ortolani}, {Paletou}, {Palou{\v{s}}}, {Pickering}, {Quirrenbach}, {Re Fiorentin}, {Read}, {Romano}, {Ryde}, {Sanna}, {Santos}, {Seabroke}, {Spina}, {Steinmetz}, {Stonkut{\'e}}, {Sutorius}, {Th{\'e}venin}, {Tosi}, {Tsantaki},
  {Wright}, {Wyse}, {Zoccali}, {Zorec}, \& {Zucker}}]{Gilmore2022}
{Gilmore}, G., {Randich}, S., {Worley}, C.~C., {et~al.} 2022, \aap, 666, A120

\bibitem[{{Grand} {et~al.}(2012){Grand}, {Kawata}, \& {Cropper}}]{Grand2012}
{Grand}, R. J.~J., {Kawata}, D., \& {Cropper}, M. 2012, \mnras, 421, 1529

\bibitem[{{Higgins} {et~al.}(2023){Higgins}, {Vink}, {Hirschi}, {Laird}, \& {Sabhahit}}]{Higgins2023MNRAS.526..534H}
{Higgins}, E.~R., {Vink}, J.~S., {Hirschi}, R., {Laird}, A.~M., \& {Sabhahit}, G.~N. 2023, \mnras, 526, 534

\bibitem[{{Hou}(2021)}]{Hou2021}
{Hou}, L.~G. 2021, Frontiers in Astronomy and Space Sciences, 8, 103

\bibitem[{{Hourihane} {et~al.}(2023){Hourihane}, {Fran{\c{c}}ois}, {Worley}, {Magrini}, {Gonneau}, {Casey}, {Gilmore}, {Randich}, {Sacco}, {Recio-Blanco}, {Korn}, {Allende Prieto}, {Smiljanic}, {Blomme}, {Bragaglia}, {Walton}, {Van Eck}, {Bensby}, {Lanzafame}, {Frasca}, {Franciosini}, {Damiani}, {Lind}, {Bergemann}, {Bonifacio}, {Hill}, {Lobel}, {Montes}, {Feuillet}, {Tautvai{\v{s}}ien{\.{e}}}, {Guiglion}, {Tabernero}, {Gonz{\'a}lez Hern{\'a}ndez}, {Gebran}, {Van der Swaelmen}, {Mikolaitis}, {Daflon}, {Merle}, {Morel}, {Lewis}, {Gonz{\'a}lez Solares}, {Murphy}, {Jeffries}, {Jackson}, {Feltzing}, {Prusti}, {Carraro}, {Biazzo}, {Prisinzano}, {Jofr{\'e}}, {Zaggia}, {Drazdauskas}, {Stonkut{\'e}}, {Marfil}, {Jim{\'e}nez-Esteban}, {Mahy}, {Guti{\'e}rrez Albarr{\'a}n}, {Berlanas}, {Santos}, {Morbidelli}, {Spina}, \& {Minkevi{\v{c}}i{\={u}}t{\.{e}}}}]{Hourihane2023A&A...676A.129H}
{Hourihane}, A., {Fran{\c{c}}ois}, P., {Worley}, C.~C., {et~al.} 2023, \aap, 676, A129

\bibitem[{{Hunt} {et~al.}(2019){Hunt}, {Bub}, {Bovy}, {Mackereth}, {Trick}, \& {Kawata}}]{Hunt2019}
{Hunt}, J. A.~S., {Bub}, M.~W., {Bovy}, J., {et~al.} 2019, \mnras, 490, 1026

\bibitem[{Hunter(2007)}]{Hunter2007}
Hunter, J.~D. 2007, Computing in Science \& Engineering, 9, 90

\bibitem[{{Khoperskov} \& {Gerhard}(2022)}]{Khoperskov_2022}
{Khoperskov}, S. \& {Gerhard}, O. 2022, \aap, 663, A38

\bibitem[{{La Vigne} {et~al.}(2006){La Vigne}, {Vogel}, \& {Ostriker}}]{LaVigne2006}
{La Vigne}, M.~A., {Vogel}, S.~N., \& {Ostriker}, E.~C. 2006, \apj, 650, 818

\bibitem[{{Lemasle} {et~al.}(2022){Lemasle}, {Lala}, {Kovtyukh}, {Hanke}, {Prudil}, {Bono}, {Braga}, {da Silva}, {Fabrizio}, {Fiorentino}, {Fran{\c{c}}ois}, {Grebel}, \& {Kniazev}}]{Lemasle2022}
{Lemasle}, B., {Lala}, H.~N., {Kovtyukh}, V., {et~al.} 2022, \aap, 668, A40

\bibitem[{{L{\'e}pine} {et~al.}(2011){L{\'e}pine}, {Cruz}, {Scarano}, {Barros}, {Dias}, {Pomp{\'e}ia}, {Andrievsky}, {Carraro}, \& {Famaey}}]{Lepine2011}
{L{\'e}pine}, J.~R.~D., {Cruz}, P., {Scarano}, Jr., S., {et~al.} 2011, \mnras, 417, 698

\bibitem[{{Lin} {et~al.}(2022){Lin}, {Xu}, {Hou}, {Liu}, {Li}, {Hao}, {Li}, \& {Bian}}]{Lin_2022}
{Lin}, Z., {Xu}, Y., {Hou}, L., {et~al.} 2022, \apj, 931, 72

\bibitem[{{Liu} {et~al.}(2025){Liu}, {He}, {Luo}, \& {Wang}}]{Liu2025}
{Liu}, X., {He}, Z., {Luo}, Y., \& {Wang}, K. 2025, arXiv e-prints, arXiv:2501.14215

\bibitem[{{Magrini} {et~al.}(2023){Magrini}, {Viscasillas V{\'a}zquez}, {Spina}, {Randich}, {Romano}, {Franciosini}, {Recio-Blanco}, {Nordlander}, {D'Orazi}, {Baratella}, {Smiljanic}, {Dantas}, {Pasquini}, {Spitoni}, {Casali}, {Van der Swaelmen}, {Bensby}, {Stonkute}, {Feltzing}, {Sacco}, {Bragaglia}, {Pancino}, {Heiter}, {Biazzo}, {Gilmore}, {Bergemann}, {Tautvai{\v{s}}ien{\.{e}}}, {Worley}, {Hourihane}, {Gonneau}, \& {Morbidelli}}]{Magrini2023}
{Magrini}, L., {Viscasillas V{\'a}zquez}, C., {Spina}, L., {et~al.} 2023, \aap, 669, A119

\bibitem[{{Minchev} {et~al.}(2012){Minchev}, {Famaey}, {Quillen}, {Di Matteo}, {Combes}, {Vlaji{\'c}}, {Erwin}, \& {Bland -Hawthorn}}]{Minchev2012}
{Minchev}, I., {Famaey}, B., {Quillen}, A.~C., {et~al.} 2012, \aap, 548, A126

\bibitem[{{Nomoto} {et~al.}(2013){Nomoto}, {Kobayashi}, \& {Tominaga}}]{Nomoto2013ARA&A..51..457N}
{Nomoto}, K., {Kobayashi}, C., \& {Tominaga}, N. 2013, \araa, 51, 457

\bibitem[{{Palicio} {et~al.}(2023){Palicio}, {Recio-Blanco}, {Poggio}, {Antoja}, {McMillan}, \& {Spitoni}}]{Palicio2023}
{Palicio}, P.~A., {Recio-Blanco}, A., {Poggio}, E., {et~al.} 2023, \aap, 670, L7

\bibitem[{pandas~development team(2020)}]{pandas2020}
pandas~development team, T. 2020, pandas-dev/pandas: Pandas

\bibitem[{{Poggio} {et~al.}(2021){Poggio}, {Drimmel}, {Cantat-Gaudin}, {Ramos}, {Ripepi}, {Zari}, {Andrae}, {Blomme}, {Chemin}, {Clementini}, {Figueras}, {Fouesneau}, {Fr{\'e}mat}, {Lobel}, {Marshall}, {Muraveva}, \& {Romero-G{\'o}mez}}]{Poggio2021}
{Poggio}, E., {Drimmel}, R., {Cantat-Gaudin}, T., {et~al.} 2021, \aap, 651, A104

\bibitem[{{Poggio} {et~al.}(2022){Poggio}, {Recio-Blanco}, {Palicio}, {Re Fiorentin}, {de Laverny}, {Drimmel}, {Kordopatis}, {Lattanzi}, {Schultheis}, {Spagna}, \& {Spitoni}}]{Poggio2022}
{Poggio}, E., {Recio-Blanco}, A., {Palicio}, P.~A., {et~al.} 2022, \aap, 666, L4

\bibitem[{Pérez-Villegas {et~al.}(2015)Pérez-Villegas, Gómez, \& Pichardo}]{Perez_Villegas_2015}
Pérez-Villegas, A., Gómez, G.~C., \& Pichardo, B. 2015, Monthly Notices of the Royal Astronomical Society, 451, 2922

\bibitem[{{Ram{\'\i}rez} {et~al.}(2014){Ram{\'\i}rez}, {Mel{\'e}ndez}, {Bean}, {Asplund}, {Bedell}, {Monroe}, {Casagrande}, {Schirbel}, {Dreizler}, {Teske}, {Tucci Maia}, {Alves-Brito}, \& {Baumann}}]{ramirez2014A&A...572A..48R}
{Ram{\'\i}rez}, I., {Mel{\'e}ndez}, J., {Bean}, J., {et~al.} 2014, \aap, 572, A48

\bibitem[{{Randich} {et~al.}(2022){Randich}, {Gilmore}, {Magrini}, {Sacco}, {Jackson}, {Jeffries}, {Worley}, {Hourihane}, {Gonneau}, {Viscasillas Vazquez}, {Franciosini}, {Lewis}, {Alfaro}, {Allende Prieto}, {Bensby}, {Blomme}, {Bragaglia}, {Flaccomio}, {Fran{\c{c}}ois}, {Irwin}, {Koposov}, {Korn}, {Lanzafame}, {Pancino}, {Recio-Blanco}, {Smiljanic}, {Van Eck}, {Zwitter}, {Asplund}, {Bonifacio}, {Feltzing}, {Binney}, {Drew}, {Ferguson}, {Micela}, {Negueruela}, {Prusti}, {Rix}, {Vallenari}, {Bayo}, {Bergemann}, {Biazzo}, {Carraro}, {Casey}, {Damiani}, {Frasca}, {Heiter}, {Hill}, {Jofr{\'e}}, {de Laverny}, {Lind}, {Marconi}, {Martayan}, {Masseron}, {Monaco}, {Morbidelli}, {Prisinzano}, {Sbordone}, {Sousa}, {Zaggia}, {Adibekyan}, {Bonito}, {Caffau}, {Daflon}, {Feuillet}, {Gebran}, {Gonzalez Hernandez}, {Guiglion}, {Herrero}, {Lobel}, {Maiz Apellaniz}, {Merle}, {Mikolaitis}, {Montes}, {Morel}, {Soubiran}, {Spina}, {Tabernero}, {Tautvai{\v{s}}iene}, {Traven}, {Valentini}, {Van der Swaelmen}, {Villanova}, {Wright},
  {Abbas}, {Aguirre B{\o}rsen-Koch}, {Alves}, {Balaguer-Nunez}, {Barklem}, {Barrado}, {Berlanas}, {Binks}, {Bressan}, {Capuzzo-Dolcetta}, {Casagrande}, {Casamiquela}, {Collins}, {D'Orazi}, {Dantas}, {Debattista}, {Delgado-Mena}, {Di Marcantonio}, {Drazdauskas}, {Evans}, {Famaey}, {Franchini}, {Fr{\'e}mat}, {Friel}, {Fu}, {Geisler}, {Gerhard}, {Gonzalez Solares}, {Grebel}, {Gutierrez Albarran}, {Hatzidimitriou}, {Held}, {Jim{\'e}nez-Esteban}, {J{\"o}nsson}, {Jordi}, {Khachaturyants}, {Kordopatis}, {Kos}, {Lagarde}, {Mahy}, {Mapelli}, {Marfil}, {Martell}, {Messina}, {Miglio}, {Minchev}, {Moitinho}, {Montalban}, {Monteiro}, {Morossi}, {Mowlavi}, {Mucciarelli}, {Murphy}, {Nardetto}, {Ortolani}, {Paletou}, {Palou{\v{s}}}, {Paunzen}, {Pickering}, {Quirrenbach}, {Re Fiorentin}, {Read}, {Romano}, {Ryde}, {Sanna}, {Santos}, {Seabroke}, {Spagna}, {Steinmetz}, {Stonkut{\'e}}, {Sutorius}, {Th{\'e}venin}, {Tosi}, {Tsantaki}, {Vink}, {Wright}, {Wyse}, {Zoccali}, {Zorec}, {Zucker}, \& {Walton}}]{Randich2022}
{Randich}, S., {Gilmore}, G., {Magrini}, L., {et~al.} 2022, \aap, 666, A121

\bibitem[{{Recio-Blanco} {et~al.}(2023){Recio-Blanco}, {de Laverny}, {Palicio}, {Kordopatis}, {{\'A}lvarez}, {Schultheis}, {Contursi}, {Zhao}, {Torralba Elipe}, {Ordenovic}, {Manteiga}, {Dafonte}, {Oreshina-Slezak}, {Bijaoui}, {Fr{\'e}mat}, {Seabroke}, {Pailler}, {Spitoni}, {Poggio}, {Creevey}, {Abreu Aramburu}, {Accart}, {Andrae}, {Bailer-Jones}, {Bellas-Velidis}, {Brouillet}, {Brugaletta}, {Burlacu}, {Carballo}, {Casamiquela}, {Chiavassa}, {Cooper}, {Dapergolas}, {Delchambre}, {Dharmawardena}, {Drimmel}, {Edvardsson}, {Fouesneau}, {Garabato}, {Garc{\'\i}a-Lario}, {Garc{\'\i}a-Torres}, {Gavel}, {Gomez}, {Gonz{\'a}lez-Santamar{\'\i}a}, {Hatzidimitriou}, {Heiter}, {Jean-Antoine Piccolo}, {Kontizas}, {Korn}, {Lanzafame}, {Lebreton}, {Le Fustec}, {Licata}, {Lindstr{\o}m}, {Livanou}, {Lobel}, {Lorca}, {Magdaleno Romeo}, {Marocco}, {Marshall}, {Mary}, {Nicolas}, {Pallas-Quintela}, {Panem}, {Pichon}, {Riclet}, {Robin}, {Rybizki}, {Santove{\~n}a}, {Silvelo}, {Smart}, {Sarro}, {Sordo}, {Soubiran}, {S{\"u}veges},
  {Ulla}, {Vallenari}, {Zorec}, {Utrilla}, \& {Bakker}}]{RecioBlanco2023}
{Recio-Blanco}, A., {de Laverny}, P., {Palicio}, P.~A., {et~al.} 2023, \aap, 674, A29

\bibitem[{{Reid} {et~al.}(2019){Reid}, {Menten}, {Brunthaler}, {Zheng}, {Dame}, {Xu}, {Li}, {Sakai}, {Wu}, {Immer}, {Zhang}, {Sanna}, {Moscadelli}, {Rygl}, {Bartkiewicz}, {Hu}, {Quiroga-Nu{\~n}ez}, \& {van Langevelde}}]{Reid2019}
{Reid}, M.~J., {Menten}, K.~M., {Brunthaler}, A., {et~al.} 2019, \apj, 885, 131

\bibitem[{{Rezaei Kh.} {et~al.}(2018){Rezaei Kh.}, {Bailer-Jones}, {Hogg}, \& {Schultheis}}]{Rezaei_2018}
{Rezaei Kh.}, S., {Bailer-Jones}, C. A.~L., {Hogg}, D.~W., \& {Schultheis}, M. 2018, \aap, 618, A168

\bibitem[{{Spitoni} {et~al.}(2019){Spitoni}, {Cescutti}, {Minchev}, {Matteucci}, {Silva Aguirre}, {Martig}, {Bono}, \& {Chiappini}}]{Spitoni2019}
{Spitoni}, E., {Cescutti}, G., {Minchev}, I., {et~al.} 2019, \aap, 628, A38

\bibitem[{{Spitoni} {et~al.}(2023){Spitoni}, {Cescutti}, {Recio-Blanco}, {Minchev}, {Poggio}, {Palicio}, {Matteucci}, {Peirani}, {Barbillon}, \& {Vasini}}]{Spitoni2023}
{Spitoni}, E., {Cescutti}, G., {Recio-Blanco}, A., {et~al.} 2023, \aap, 680, A85

\bibitem[{Sánchez-Menguiano {et~al.}(2020)Sánchez-Menguiano, Sánchez, Pérez, Ruiz-Lara, Galbany, Anderson, \& Kuncarayakti}]{Sanchez_Menguiano_2020}
Sánchez-Menguiano, L., Sánchez, S.~F., Pérez, I., {et~al.} 2020, Monthly Notices of the Royal Astronomical Society, 492, 4149–4163

\bibitem[{{Tagger} {et~al.}(1987){Tagger}, {Sygnet}, {Athanassoula}, \& {Pellat}}]{Tagger1987}
{Tagger}, M., {Sygnet}, J.~F., {Athanassoula}, E., \& {Pellat}, R. 1987, \apjl, 318, L43

\bibitem[{{Taylor}(2005)}]{Taylor2005}
{Taylor}, M.~B. 2005, in Astronomical Society of the Pacific Conference Series, Vol. 347, Astronomical Data Analysis Software and Systems XIV, ed. P.~{Shopbell}, M.~{Britton}, \& R.~{Ebert}, 29

\bibitem[{{Vall{\'e}e}(2014)}]{Valle2014}
{Vall{\'e}e}, J.~P. 2014, \aj, 148, 5

\bibitem[{{Vall{\'e}e}(2016)}]{Valle2016}
{Vall{\'e}e}, J.~P. 2016, \aj, 151, 55

\bibitem[{{Van der Swaelmen} {et~al.}(2024){Van der Swaelmen}, {Viscasillas V{\'a}zquez}, {Magrini}, {Recio-Blanco}, {Palicio}, {Worley}, {Vallenari}, {Spina}, {Fran{\c{c}}ois}, {Tautvai{\v{s}}ien{\.{e}}}, {Sacco}, {Randich}, \& {de Laverny}}]{VanderSwaelmen2024}
{Van der Swaelmen}, M., {Viscasillas V{\'a}zquez}, C., {Magrini}, L., {et~al.} 2024, \aap, 690, A276

\bibitem[{{Vasini} {et~al.}(2025){Vasini}, {Spitoni}, {Matteucci}, {Cescutti}, \& {Della Valle}}]{Vasini2025}
{Vasini}, A., {Spitoni}, E., {Matteucci}, F., {Cescutti}, G., \& {Della Valle}, M. 2025, \aap, 693, A37

\bibitem[{{Veena} {et~al.}(2021){Veena}, {Schilke}, {S{\'a}nchez-Monge}, {Sormani}, {Klessen}, {Schuller}, {Colombo}, {Csengeri}, {Mattern}, \& {Urquhart}}]{Veena2021}
{Veena}, V.~S., {Schilke}, P., {S{\'a}nchez-Monge}, {\'A}., {et~al.} 2021, \apjl, 921, L42

\bibitem[{Virtanen {et~al.}(2020)Virtanen, Gommers, Oliphant, Haberland, Reddy, Cournapeau, Burovski, Peterson, Weckesser, Bright, {van der Walt}, Brett, Wilson, Millman, Mayorov, Nelson, Jones, Kern, Larson, Carey, Polat, Feng, Moore, {VanderPlas}, Laxalde, Perktold, Cimrman, Henriksen, Quintero, Harris, Archibald, Ribeiro, Pedregosa, {van Mulbregt}, \& {SciPy 1.0 Contributors}}]{Virtanen2020}
Virtanen, P., Gommers, R., Oliphant, T.~E., {et~al.} 2020, Nature Methods, 17, 261

\bibitem[{{Viscasillas V{\'a}zquez} {et~al.}(2022){Viscasillas V{\'a}zquez}, {Magrini}, {Casali}, {Tautvai{\v{s}}ien{\.{e}}}, {Spina}, {Van der Swaelmen}, {Randich}, {Bensby}, {Bragaglia}, {Friel}, {Feltzing}, {Sacco}, {Turchi}, {Jim{\'e}nez-Esteban}, {D'Orazi}, {Delgado-Mena}, {Mikolaitis}, {Drazdauskas}, {Minkevi{\v{c}}i{\={u}}t{\.{e}}}, {Stonkut{\.{e}}}, {Bagdonas}, {Montes}, {Guiglion}, {Baratella}, {Tabernero}, {Gilmore}, {Alfaro}, {Francois}, {Korn}, {Smiljanic}, {Bergemann}, {Franciosini}, {Gonneau}, {Hourihane}, {Worley}, \& {Zaggia}}]{Viscasillas2022}
{Viscasillas V{\'a}zquez}, C., {Magrini}, L., {Casali}, G., {et~al.} 2022, \aap, 660, A135

\bibitem[{{Viscasillas V{\'a}zquez} {et~al.}(2024){Viscasillas V{\'a}zquez}, {Magrini}, {Miret-Roig}, {Wright}, {Alves}, {Spina}, {Church}, {Tautvai{\v{s}}ien{\.{e}}}, \& {Randich}}]{Viscasillas2024}
{Viscasillas V{\'a}zquez}, C., {Magrini}, L., {Miret-Roig}, N., {et~al.} 2024, \aap, 689, A268

\bibitem[{{Viscasillas V{\'a}zquez} {et~al.}(2023){Viscasillas V{\'a}zquez}, {Magrini}, {Spina}, {Tautvai{\v{s}}ien{\.{e}}}, {Van der Swaelmen}, {Randich}, \& {Sacco}}]{Viscasillas2023}
{Viscasillas V{\'a}zquez}, C., {Magrini}, L., {Spina}, L., {et~al.} 2023, \aap, 679, A122

\bibitem[{{Vislosky} {et~al.}(2024){Vislosky}, {Minchev}, {Khoperskov}, {Martig}, {Buck}, {Hilmi}, {Ratcliffe}, {Bland-Hawthorn}, {Quillen}, {Steinmetz}, \& {de Jong}}]{Vislosky2024}
{Vislosky}, E., {Minchev}, I., {Khoperskov}, S., {et~al.} 2024, \mnras, 528, 3576

\bibitem[{Waskom(2021)}]{Waskom2021}
Waskom, M.~L. 2021, Journal of Open Source Software, 6, 3021

\bibitem[{{Widmark} \& {Naik}(2024)}]{Widmark_2024}
{Widmark}, A. \& {Naik}, A.~P. 2024, \aap, 686, A70

\bibitem[{{Xu} {et~al.}(2023){Xu}, {Hao}, {Liu}, {Lin}, {Bian}, {Hou}, {Li}, \& {Li}}]{Xu2023}
{Xu}, Y., {Hao}, C.~J., {Liu}, D.~J., {et~al.} 2023, \apj, 947, 54

\end{thebibliography}

\end{document}